\documentclass[11pt]{article}
\usepackage{enumerate,amsmath,amsfonts,amssymb,graphicx,dsfont}
\usepackage[english]{babel}
\usepackage[latin1]{inputenc}
\usepackage{latexsym}
\usepackage{graphicx}
\usepackage{listings}
\usepackage{subfigure}
\usepackage{pdfpages}
\usepackage{bm}
\usepackage{amsthm}
\usepackage{hyperref}
\usepackage[margin=1in]{geometry}    
\usepackage{abstract}            		
\usepackage{titlesec}
\usepackage{float}

\usepackage[section]{placeins} 

\usepackage[round]{natbib} 

\allowdisplaybreaks 

\DeclareMathOperator{\Tr}{Tr}

\title{\textbf{
				Checking Validity of Monotone Domain Mean Estimators }}

\author{
	\large
	\textsc{Cristian Oliva\thanks{coliva@colostate.edu} , Mary C. Meyer\thanks{meyer@stat.colostate.edu} and Jean D. Opsomer\thanks{jopsomer@stat.colostate.edu}} \\ [2mm] 
	\normalsize Department of Statistics, Colorado State University, Fort Collins, Colorado, USA, 80523 \\ 
}

\date{ \vspace{3mm} Updated version by \today}

\begin{document}

\maketitle

\begin{abstract}
	Estimates of population characteristics such as domain means are often expected to follow monotonicity assumptions. Recently, a method to adaptively pool neighboring domains was proposed, which ensures that the resulting domain mean estimates follow monotone constraints. The method leads to asymptotically valid estimation and inference, and can lead to substantial improvements in efficiency, in comparison with unconstrained domain estimators. However, assuming incorrect shape constraints could lead to biased estimators. Here, we develop the Cone Information Criterion for Survey Data (CIC$_s$) as a diagnostic method to measure monotonicity departures on population domain means.  We show that the criterion leads to a consistent methodology that makes an asymptotically correct decision choosing between unconstrained and constrained domain mean estimators. 
\end{abstract}
	
\section{Introduction}
Monotone population characteristics arise naturally in many survey problems. For example, average salary might be increasing in pay grade, average cholesterol level could be decreasing in physical activity time, etc. In  large-scale surveys, there is often interest in estimating the characteristics of domains within the overall population, including those of domains with small sample sizes. One possibility to handle small domains is to apply small area estimation methods. However, that requires switching from the design-based to a model-based paradigm, which can be undesirable. An alternative approach is to remain within the design-based paradigm but take advantage of qualitative assumptions about the population structure, when such are available.

Isotonic regression has been widely studied outside of the survey context. Some remarkable works on this topic include \cite{brunk55}, \cite{vaneeden56}, \cite{brunk58}, \cite{robertson88}, and \cite{silvapulle05}. In contrast, merging isotonic regression techniques into survey estimation and inference has just been studied recently. \cite{wu16} considered the case when both sampling design and monotone restrictions are taking into account on the domain estimation. They proposed a design-weighted constrained estimator by combining domain estimation and the Pooled Adjacent Violators Algorithm (PAVA) \citep{robertson88}. Further, they showed that their proposed constrained estimator improved estimation and variability of domain means, under both linearization-based and replication-based variance estimation. 

Although the constrained estimator proposed by \cite{wu16} improves the precision of the usual survey sampling estimators, it has to be used carefully since invalid population constraint assumptions could lead to biased domain mean estimators. The main objective of this work is to develop diagnostic methods to detect population departures from monotone assumptions. Particularly, we propose the Cone Information Criterion for Survey Data (CIC$_s$) as a data-driven method to determine whether or not it is better to use the constrained estimator to estimate the population domain means. The Cone Information Criterion (CIC) was originally developed for the i.i.d. \negthinspace case by \cite{meyer13}.

In Section 2, we describe the constrained estimator proposed by \citet{wu16} and explain some of its properties such as adaptive pooling domain and linearization-based variance estimation. Section 3 contains the proposed CIC$_s$ along with some of its theoretical properties. In particular, we show that CIC$_s$ is consistently choosing the correct estimator based on the underlying shape of the population domain means, in the sense that with probability going to 1 as the sample size increases, CIC$_s$ will determine that pooling of domains that violate monotonicity constraints is unwarranted. Section 4 demonstrates the performance of the CIC$_s$ under a broad variety of simulation scenarios. In Section 5 we apply our CIC$_s$ methodology to the 2011-2012 National Health and Nutrition Examination Survey (NHANES) laboratory data. Lastly, Section 6 states some general conclusions of the work developed in this paper, and contains a brief discussion about future related areas of research.

\section{Constrained Domain Mean Estimator for Survey Data}

We begin by reviewing the survey setting and the constrained estimator proposed by \citet{wu16}.  Consider a finite population $U_N=\{1,2,\dots,N\}$, and let $U_{d,N}$ denote a domain for $d=1,\dots, D$. Assume that $\{U_{d,N}\; ; \;  d=1,\dots,D \}$ constitute a partition of the population $U_N$. Denote $N_{d}$ as the population size of domain $U_{d,N}$. Given a study variable $y$, let $\overline{y}_{U_{d}}$ be the population domain means, 
\begin{equation*}
\overline{y}_{U_{d}}=\frac{\sum_{k \in U_{d,N}} y_k}{N_{d}}, \; \; \; d=1,\dots, D.
\end{equation*}
Suppose we draw a sample $s_N \subset U_N$ using the probability sampling design $p_N(\cdot)$. Let $n_N$ be the sample size of $s_N$. We are going to consider the case where the sampling design is measurable, i.e., both first-order $\pi_k=\mathbb{E}(I_k)$ and second-order $\pi_{kl}=\mathbb{E}(I_kI_l)$ inclusion probabilities are strictly positive, where $I_k$ is the indicator variable of whether $k\in s_N$ or not. Denote $s_{d,N}$ as the corresponding sample in domain $d$ obtained from $s_N$. Further, let $n_{d,N}=|s_{d,N}|$. For simplicity in our notation, we will omit the subscript $N$ from these and related quantities from now on.


Consider the problem of estimating the population domain means $\overline{y}_{U_d}$. When no qualitative information is assumed on the population domains, we can consider either the Horvitz-Thompson estimator $\widehat{y}_{s_d}$ \citep{horvitz52} or the frequently preferred H\'{a}jek estimator $\tilde{y}_{s_{d}}$ \citep{hajek71}, which are given by
\begin{equation} \label{eq:1}
\widehat{y}_{s_d}=\frac{\sum_{k \in s_d}y_k/\pi_k }{N_d}, \; \; \; \; \tilde{y}_{s_d}=\frac{\sum_{k \in s_d}y_k/\pi_k }{\widehat{N}_d},
\end{equation}
respectively, where $\widehat{N}_d=\sum_{k \in s_d}1/\pi_k$. We will refer to them as unconstrained estimators of $\overline{y}_{U_d}$. Note that both estimators in Equation \ref{eq:1} consider only the information contained in domain $d$, leading to large standard errors on domains with small sample sizes. 


Suppose now that we want to include monotonicity assumptions into the estimation stage of domain means. For instance, assume the population domain means are isotonic over the $D$ domains. That is, $\overline{y}_{U_1}\leq \overline{y}_{U_2} \leq  \dots \leq \overline{y}_{U_D}$ (analogously, $\overline{y}_{U_1} \geq \overline{y}_{U_2} \geq  \dots \geq \overline{y}_{U_D}$, but which we will not further consider explicitly here). \citet{wu16} proposed a domain mean estimator that respect monotone constraints, given by the ordered vector $\boldsymbol{\tilde{\theta}}_s=( \tilde{\theta}_{s_1}, \tilde{\theta}_{s_2}, \dots, \tilde{\theta}_{s_D} )^{\top}$ which optimizes 
\begin{equation} \label{eq:2}
\underset{\theta_1, \theta_2, \dots, \theta_D}{\min} \; \sum_{d=1}^{D} \widehat{N}_d (\tilde{y}_{s_d}-\theta_d)^2, \; \; \;  \text{ subject to } \; \; \; \theta_1 \leq \theta_2 \leq \dots \leq \theta_D.
\end{equation}
The objective function in Equation \ref{eq:2} can be written in matrix terms as $(\boldsymbol{\tilde{y}}_s-\boldsymbol{\theta})^{\top}\boldsymbol{W}_s(\boldsymbol{\tilde{y}}_s-\boldsymbol{\theta})$, where $\boldsymbol{\tilde{y}}_{s}=(\tilde{y}_{s_1},\tilde{y}_{s_2},\dots, \tilde{y}_{s_D})^{\top}$, $\boldsymbol{\theta}=(\theta_1, \theta_2, \dots, \theta_D)^{\top}$, $\boldsymbol{W}_s=\text{diag}(\widehat{N}_1/\widehat{N},\widehat{N}_2/\widehat{N}, \dots, \widehat{N}_D/\widehat{N})$ is a consistent estimator of $\boldsymbol{W}_U=\text{diag}(N_1/N, N_2/N,\dots, N_D/N)$, and $\widehat{N}=\sum_{d=1}^{D}\widehat{N}_d$.

Following \citet{brunk55}, the general closed form solution for the constrained problem in Equation \ref{eq:2} can be expressed as the set of pooled weighted domain means given by
\begin{equation} \label{eq:3}
\tilde{\theta}_{s_d}=\underset{i \leq d}{\max} \; \underset{d \leq j}{\min} \; \tilde{y}_{s_{i:j}}, \; \; \; \text{ where } \; \; \; \tilde{y}_{s_{i:j}}=\frac{\sum_{d=i}^{j}\widehat{N}_d \tilde{y}_{s_d} }{\sum_{d=i}^{j} \widehat{N}_d}=\frac{\sum_{k \in s_{i:j}} y_k/\pi_k }{\sum_{k \in s_{i:j}} 1/\pi_k}, 
\end{equation} 
where $s_{i:j}=s_i\cup \dots \cup s_j$ for $1 \leq i\leq j \leq D$. Moreover, we can make use of the Pooled Adjacent Violator Algorithm PAVA \citep{robertson88} along with $\boldsymbol{\tilde{y}}_s$ and the weights $\widehat{N}_1, \widehat{N}_2, \dots, \widehat{N}_D$ to compute efficiently the constrained estimator $\boldsymbol{\tilde{\theta}}_s$. Observe that the constrained estimator in Equation \ref{eq:3} consists of adaptively collapsing neighboring domains. Furthermore, the above procedure can be simplified in the obvious way when applied to the Horvitz-Thompson estimator  $\boldsymbol{\widehat{y}}_{s}=(\widehat{y}_{s_1},\widehat{y}_{s_2},\dots, \widehat{y}_{s_D})^{\top}$ with weights $N_1, N_2, \dots, N_D$, leading to the constrained estimator vector $\boldsymbol{\widehat{\theta}}_s$ with entries of the form $\widehat{y}_{s_{i:j}}$.  We refer to \citet{wu16} for a discussion of the properties of these constrained estimators, including design consistency and asymptotic distribution.

We conclude this section by defining some of the quantities we will use in the development of the CIC$_s$. Note that the estimator $\boldsymbol{\widehat{\theta}}_s$ has a random weighted projection matrix $\boldsymbol{\widehat{P}}_s$ associated with it, which is defined by the pooling obtained from the PAVA and the weights $N_1, N_2,\dots, N_D$. That is, $\boldsymbol{\widehat{P}}_s$ is the matrix such that $\boldsymbol{\widehat{\theta}}_s=\boldsymbol{\widehat{P}}_s \boldsymbol{\widehat{y}}_s$. For example, suppose $D=3$ and that PAVA chooses to pool domains 1 and 2, but not to pool domain 2 and 3. Hence, $\widehat{\theta}_{s_1}=\widehat{\theta}_{s_2}=(N_1 \widehat{y}_{s_1}+N_2 \widehat{y}_{s_2})/(N_1+N_2)$, and $\widehat{\theta}_{s_3}=\widehat{y}_{s_3}$. Then, 
\begin{equation*}
\boldsymbol{\widehat{P}}_s=\left(\begin{array}{ccc}
\frac{N_1}{N_1+N_2} & \frac{N_2}{N_1+N_2} & 0 \\
\frac{N_1}{N_1+N_2} & \frac{N_2}{N_1+N_2} & 0 \\
0 & 0 & 1
\end{array}\right).
\end{equation*}
Let $\boldsymbol{\widehat{\Sigma}}=\{\widehat{\Sigma}_{ij}\}$ be the unbiased estimator of the covariance matrix of $\boldsymbol{\widehat{y}}_{s}$, given by
\begin{equation*}
\widehat{\Sigma}_{ij}=\frac{1}{N_iN_j}\sum_{k \in s_i}\sum_{l \in  s_j}\frac{\Delta_{kl}}{\pi_{kl}}\frac{y_k}{\pi_k}\frac{y_l}{\pi_l} \text{, } \; \; \; i,j=1,2,\dots,D,
\end{equation*}
where $\Delta_{kl}=\pi_{kl}-\pi_k\pi_l$. Further, for any $i\leq j$, let $\overline{y}_{U_{i:j}}$ be the pooled population mean of domains $i$ through $j$. That is,  
\begin{equation*}
\overline{y}_{U_{i:j}}=\frac{\sum_{k \in U_{i:j}}y_k}{N_{i:j}}, \; \; \; \text{ where } \; \; \; N_{i:j}=\sum_{d=i}^{j}N_d, 
\end{equation*} 
and $U_{i:j}=U_i \cup \dots \cup U_j$. 

For any indexes $i_1,i_2,j_i,j_2$ such that $i_1 \leq j_1$ and $i_2 \leq j_2$, let $\tilde{y}_{s_{i_1:j_1}}$, $\tilde{y}_{s_{i_2:j_2}}$ be the H\'ajek estimators of $\overline{y}_{U_{i_1:j_1}}$ and $\overline{y}_{U_{i_2:j_2}}$, respectively.  By standard linearization arguments \citep[Chapter 5]{sarndal92}, the approximated covariance of $\tilde{y}_{s_{i_1:j_1}}$ and $\tilde{y}_{s_{i_2:j_2}}$ is given by
\begin{equation} \label{eq:4}
AC(\tilde{y}_{s_{i_1:j_1}}, \tilde{y}_{s_{i_2:j_2}})=\frac{1}{N_{i_1:j_1} N_{i_2:j_2}}\sum_{k \in U_{i_1 :j_1}}\sum_{l \in  U_{i_2:j_2}}\Delta_{kl}\left(\frac{y_k-\overline{y}_{U_{i_1:j_1}}}{\pi_k}\right) \left(\frac{y_l-\overline{y}_{U_{i_2:j_2}}}{\pi_l}\right).
\end{equation}
Moreover, given that $\pi_{kl}>0$ for all $k,l \in U$, a design consistent estimator of the approximate covariance in Equation \ref{eq:4} is
\begin{equation} \label{eq:5}
\widehat{AC}(\tilde{y}_{s_{i_1:j_1}}, \tilde{y}_{s_{i_2:j_2}})=\frac{1}{\widehat{N}_{i_1:j_1} \widehat{N}_{i_2:j_2}}\sum_{k \in s_{i_1 :j_1}}\sum_{l \in  s_{i_2:j_2}}\frac{\Delta_{kl}}{\pi_{kl}}\left(\frac{y_k-\tilde{y}_{s_{i_1:j_1}}}{\pi_k}\right) \left(\frac{y_l-\tilde{y}_{s_{i_2:j_2}}}{\pi_l}\right),
\end{equation}
where $\widehat{N}_{i:j}=\sum_{d=i}^{j}\widehat{N}_d$. 

\section{Main results}

In this section, we present the Cone Information Criterion for Survey Data (CIC$_s$). The CIC$_s$ is a tool that may be used to validate the monotone estimator in Equation \ref{eq:2} as an appropriate estimator of population domain means. In what follows, we define the CIC$_s$ for the Horvitz-Thompson estimator and propose a natural extension that applies to the H\'ajek setting. Further, main properties of the CIC$_s$ are shown along with their theoretical foundation.  

\subsection{Cone Information Criterion for Survey Data (CIC$_s$)} 

For the Horvitz-Thompson estimator, we define the CIC$_s$ as 
\begin{equation} \label{eq:6}
\text{CIC}_\text{s}(\boldsymbol{\widehat{\theta}}_s)= (\boldsymbol{\widehat{y}}_{s}-\boldsymbol{\widehat{\theta}}_s)^{\top} \boldsymbol{W}_U (\boldsymbol{\widehat{y}}_{s}-\boldsymbol{\widehat{\theta}}_s) + 2\Tr\left( \boldsymbol{W}_U \boldsymbol{\widehat{P}}_s \boldsymbol{\widehat{\Sigma}}\right),
\end{equation}
where $\boldsymbol{\widehat{P}}_s$ is the projection matrix associated with $\boldsymbol{\widehat{\theta}}_s$.

The proposed CIC$_s$ shares similar features with the Akaike Information Criterion (AIC) \citep{akaike73} and the Bayesian Information Criterion (BIC) \citep{schwarz78}, which have been broadly used for model selection. The first term measures the deviation between the constrained estimator $\widehat{\boldsymbol{\theta}}_s$ and the unconstrained estimator $\boldsymbol{\overline{y}}_s$, while the second term can be seen as a penalty for the complexity of the constrained estimator. The penalty term is large when the number of different groups chosen by the constrained estimator is also large, meaning that the number of different parameters to estimate (or effective degrees of freedom) of the constrained estimator is high.

The development of $\text{CIC}_\text{s}$ proceeds similarly as for the Cone Information Criterion (CIC) proposed by \cite{meyer13}. Its motivation comes from properties of the Predictive Squared Error (PSE) under the Horvitz-Thompson setting, which is defined as
\begin{equation} \label{eq:7}
\text{PSE}(\boldsymbol{\widehat{\theta}}_s)=\mathbb{E}\left[ (\boldsymbol{\widehat{y}}_{s^*}-\boldsymbol{\widehat{\theta}}_{s})^{\top} \boldsymbol{W}_U (\boldsymbol{\widehat{y}}_{s^*}-\boldsymbol{\widehat{\theta}}_{s}) \right]
\end{equation} 
where $\boldsymbol{\widehat{y}}_{s^*}$ is the vector of Horvitz-Thompson domain mean estimators obtained from a sample $s^*$ that is independent to $s$, where $s^*$ is drawn using the same probability sampling design as $s$. Furthermore, define the Sum of Squared Errors (SSE) as
\begin{equation*}
\text{SSE}(\boldsymbol{\widehat{\theta}}_s)= (\boldsymbol{\widehat{y}}_{s}-\boldsymbol{\widehat{\theta}}_s)^{\top} \boldsymbol{W}_U (\boldsymbol{\widehat{y}}_{s}-\boldsymbol{\widehat{\theta}}_s). 
\end{equation*}

We define CIC$_s(\boldsymbol{\widehat{\theta}}_s)$ as an estimator of PSE$(\boldsymbol{\widehat{\theta}}_s)$ that involves SSE$(\boldsymbol{\widehat{\theta}}_s)$. Proposition 1 establishes a relationship between PSE$(\boldsymbol{\widehat{\theta}}_s)$ and SSE$(\boldsymbol{\widehat{\theta}}_s)$; its proof and all subsequent ones are included in the Appendix. \\

\textbf{Proposition 1.} $\text{PSE}(\boldsymbol{\widehat{\theta}}_s)=\mathbb{E}\left[\text{SSE}(\boldsymbol{\widehat{\theta}}_s)\right]+ 2\Tr \left[ \boldsymbol{W}_U \text{cov} ( \widehat{\boldsymbol{\theta}}_{s}, \widehat{\boldsymbol{y}}_s ) \right]$.\\

Motivated by Proposition 1, an estimate of PSE$(\boldsymbol{\widehat{\theta}}_s)$ can be derived by estimating both $\mathbb{E}\left[\text{SSE}(\boldsymbol{\widehat{\theta}}_s)\right]$ and $\text{cov} ( \widehat{\boldsymbol{\theta}}_{s}, \widehat{\boldsymbol{y}}_s )$. The first term has a straightforward unbiased estimator SSE$(\boldsymbol{\widehat{\theta}}_s)$, and an estimator for the covariance term can be obtained using the observed pooling on $\widehat{\boldsymbol{\theta}}_{s}$. As we will show later, the latter term can be estimated by the asymptotically unbiased estimator $\boldsymbol{\widehat{P}}_{s}\boldsymbol{\widehat{\Sigma}}$ under certain assumptions. That produces the proposed CIC$_s$ in Equation \ref{eq:6}.

However, recall that the use of the Horvitz-Thompson estimator requires information about the population domain sizes $N_d$, which is not frequently the case in many practical survey applications. Therefore, analogously to Equation \ref{eq:6}, we extend the CIC$_s$ to the H\'ajek setting by using the estimator $(\boldsymbol{\tilde{y}}_{s}-\boldsymbol{\tilde{\theta}}_s)^{\top} \boldsymbol{W}_s (\boldsymbol{\tilde{y}}_{s}-\boldsymbol{\tilde{\theta}}_s)$ instead of SSE$(\boldsymbol{\widehat{\theta}}_s)$, and $\boldsymbol{W}_s \widehat{\text{cov}}(\boldsymbol{\tilde{\theta}}_s, \boldsymbol{\tilde{y}}_s )$ instead of $\boldsymbol{W}_U \boldsymbol{\widehat{P}}_{s}\boldsymbol{\widehat{\Sigma}}$; where $\widehat{\text{cov}}(\boldsymbol{\tilde{\theta}}_s, \boldsymbol{\tilde{y}}_s )$ denotes the estimator of the covariance matrix of $\boldsymbol{\tilde{\theta}}_s$ and  $\boldsymbol{\tilde{y}}_s$, which is based on the observed pooling of $\boldsymbol{\tilde{\theta}}_s$ and is defined element-wise as
\begin{equation*}
\widehat{\text{cov}}(\boldsymbol{\tilde{\theta}}_s, \boldsymbol{\tilde{y}}_s )_{ij}=\widehat{AC}(\tilde{\theta}_{s_i}, \tilde{y}_{s_j}), \; \; \; \text{ for } \; \; \; i,j=1,2, \dots, D.
\end{equation*}
Hence, the proposed CIC$_s$ for the H\'ajek estimator setting is 
\begin{equation} \label{eq:8}
\text{CIC}_\text{s}(\boldsymbol{\tilde{\theta}}_s)= (\boldsymbol{\tilde{y}}_{s}-\boldsymbol{\tilde{\theta}}_s)^{\top} \boldsymbol{W}_s (\boldsymbol{\tilde{y}}_{s}-\boldsymbol{\tilde{\theta}}_s) + 2\Tr\left[ \boldsymbol{W}_s \widehat{\text{cov}}(\boldsymbol{\tilde{\theta}}_s, \boldsymbol{\tilde{y}}_s ) \right].
\end{equation}


\subsection{Assumptions}

In order to state properly our theoretical results, we need to consider some required assumptions.

\begin{itemize}
	\item[\textbf{(A1)}] The number of domains $D$ is a fixed known constant.
	\item[\textbf{(A2)}] The non-random sample size $n_N$ satisfies $0<\underset{N\rightarrow \infty}{\lim}\frac{n_N}{N}<1$. 
	\item[\textbf{(A3)}] $\underset{N \rightarrow \infty}{\limsup} \; \frac{1}{N} \underset{k \in U_N}{\sum} y_k^4 < \infty $. 
	\item[\textbf{(A4)}]  $0<\gamma_d=\underset{N\rightarrow \infty}{\lim}\frac{N_d}{N} <1 $ for $d=1,2,\dots,D$. Also, for some constants $\mu_1,\mu_2,\dots, \mu_D$ and any integers $i,j$ such that $1\leq i\leq j \leq D$, then $\overline{y}_{U_{i:j}}-\mu_{i:j}=O(N^{-1/2})$ with $\mu_{i:j}=\sum_{d=i}^{j}\gamma_d\mu_d$.
	\item[\textbf{(A5)}] For all $N$, $\underset{k \in U_N}{\min} \; \pi_k \geq \lambda >0$, $\underset{k,l \in U_N}{\min} \; \pi_{kl} \geq \lambda^*>0$, and $\underset{N\rightarrow \infty}{\limsup} \; n_N \underset{k,l \in U_N : \; k \neq l}{\max} \; |\Delta_{kl}| < \infty$. 
	\item[\textbf{(A6)}] $\underset{N \rightarrow \infty}{\lim} \; \underset{(k_1,k_2,k_3,k_4) \in D_{4,N}}{\max} \left|\mathbb{E}\left[(I_{k_1}I_{k_2}-\pi_{k_1k_2})(I_{k_3}I_{k_4}-\pi_{k_3k_4})\right]\right|=0$, where $D_{4,N}$ denotes the set of all distinct $4-$tuples $(k_1,k_2,k_3,k_4)$ from $U_N$.
	\item[\textbf{(A7)}] $\lim_{N \rightarrow \infty} \max_{(k_1,k_2,k_3) \in D_{3,N}}|\mathbb{E}[(I_{k_1}-\pi_{k_1})^2(I_{k_2}-\pi_{k_2})(I_{k_3}-\pi_{k_3}) ]|=0$.
	\item[\textbf{(A8)}] $\limsup_{N \rightarrow \infty} n_N \max_{(k_1,k_2,k_3,k_4) \in D_{4,N}}|\mathbb{E}[(I_{k_1}-\pi_{k_1})(I_{k_2}-\pi_{k_2})(I_{k_3}-\pi_{k_3})(I_{k_4}-\pi_{k_4}) ]|=0$.
\end{itemize}

Assumption (A1) states that the number of domains $D$ will not change as the population size changes. Assumption (A2) declares that the sample size is asymptotically strictly less than the population size but greater than zero, which intuitively means that the sample and the population size are of the same order. The boundedness property of the finite population fourth moment in Assumption (A3) is used several times in our proofs to show that the approximated scaled covariances in Equation \ref{eq:4} are asymptotically bounded, and also, that their estimators are consistent for them. In addition, Assumption (A4) is used to assure that the population size and the subpopulation size are of the same order. Further, it establishes that the pooled population domain means converge to some constant limiting domain means with rate $N^{-1/2}$. The consistency result of CIC$_s$ is based on whether the constants $\mu_1,\mu_2,\dots, \mu_D$ are strictly monotone or not. Assumption (A5) implies that both first and second-order inclusion probabilities can not tend to zero as $N$ increases. Moreover, this assumption states that the sampling design covariances $\Delta_{kl}$ ($k \neq l$) tend to zero, i.e., sampling designs that produces asymptotically highly correlated elements are not allowed. Lastly, Assumptions (A6)-(A8) are similar to the higher order assumptions considered by \cite{breidt2000}. These assumptions involve fourth moment conditions on the sampling design. These assumptions hold for simple random sampling without replacement and for stratified simple random sampling with fixed stratum boundaries \citep{breidt2000}.

\subsection{Properties of CIC$_s$}

Under above assumptions, CIC$_s(\boldsymbol{\widehat{\theta}}_s)$ has the property of being an asymptotically unbiased estimator of PSE$(\boldsymbol{\widehat{\theta}}_s)$ when the pooling obtained from applying the PAVA to the vector $\boldsymbol{\mu}=(\mu_1, \mu_2, \dots, \mu_D)^{\top}$ with weights $\gamma_1, \gamma_2, \dots, \gamma_D$ is unique. To show that, we first prove that there are certain poolings which are chosen with probability tending to zero as $N$ tends to infinity. This is stated in Theorem 1, which makes use of the Greatest Convex Minorant (GCM). 

\begin{figure}[ht!]
	\subfigure[Sample domain means and monotone estimates.]{\label{fig:1a} \includegraphics[width=.49\textwidth]{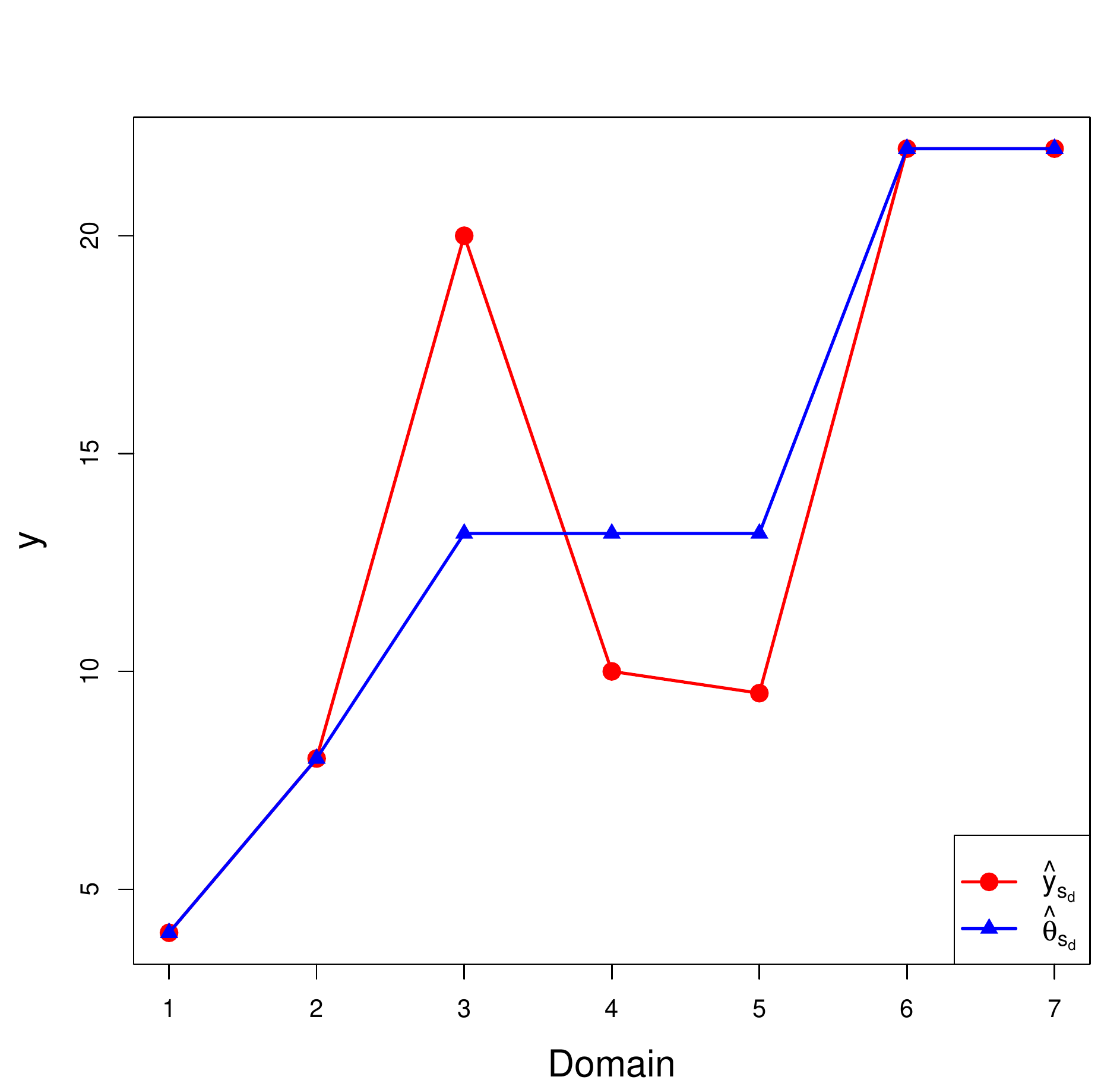}}\
	\subfigure[Cumulative sum diagram and GCM.]{\label{fig:1b} \includegraphics[width=.49\textwidth]{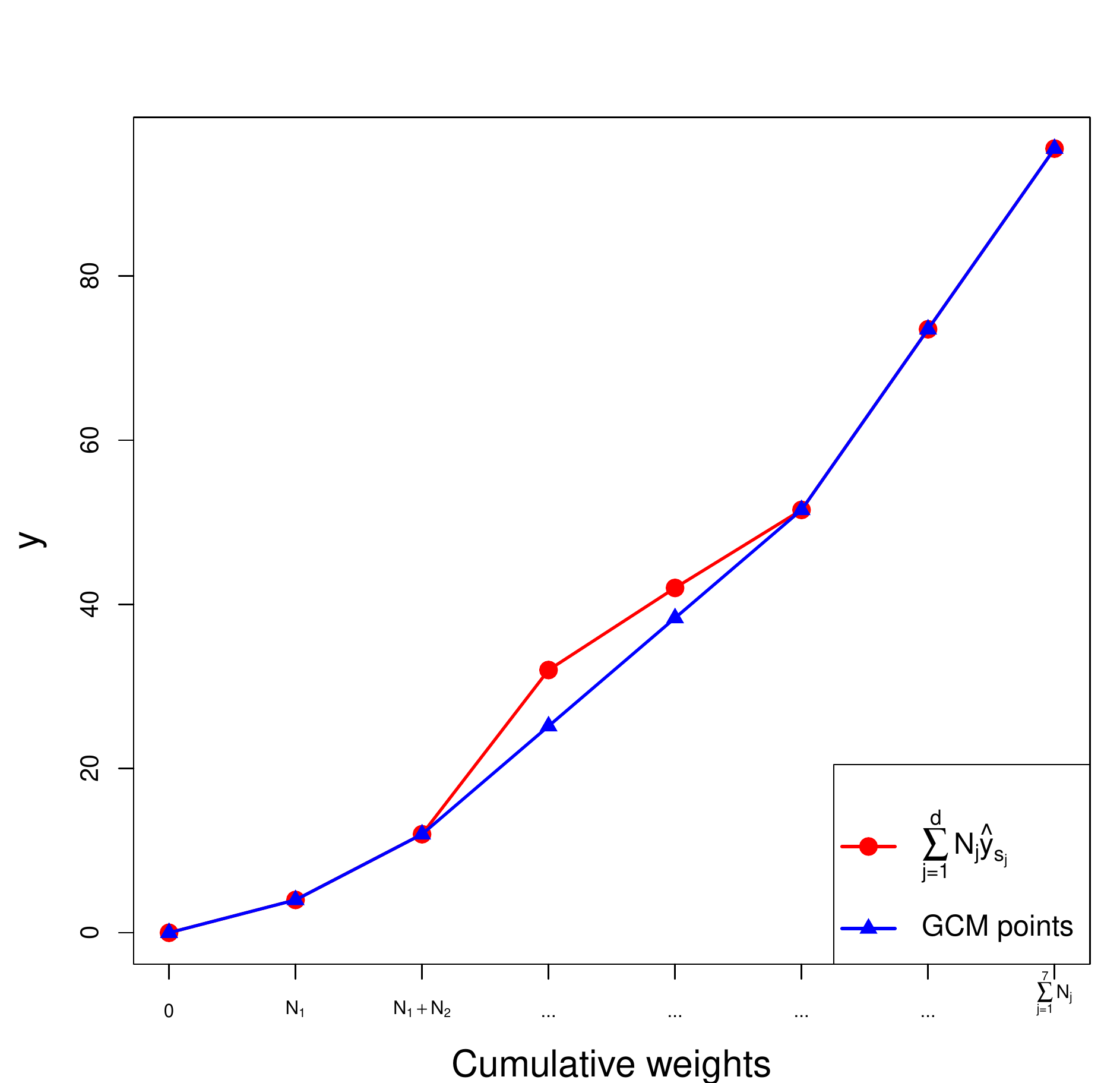}}\\
	\caption{GCM example.}
	\label{fig:1}
\end{figure}

The GCM provides of an illustrative way to express monotone estimators. Figure \ref{fig:1} displays an example of sample domain means with their respective monotone estimates (Figure \ref{fig:1a}), and a plot of their corresponding cumulative sum diagram and GCM (Figure \ref{fig:1b}). The GCM is conformed by $D+1$ points, indexed from 0 to $D$, and their left-hand slopes are the $\widehat{\theta}_{s_d}$ values. The points indexed by 0 and $D$ are the boundaries of the GCM, and the rest are its interior points. Three possible scenarios can be identified for each of the interior points: the slope of the GCM changes (\textsl{corner points}); the GCM slope does not change and the cumulative sum coincides with the minorant (\textsl{flat spots}); or the GCM slope does not change but the cumulative sum is strictly above the minorant (\textsl{points above the GCM}). The example displayed in Figure \ref{fig:1b} shows that the indexes 1, 2, 5 correspond to corner points, the index 6 to a flat spot, and the indexes 3, 4 to points above the GCM. In particular, note that flat spots correspond to cases where consecutive domain means are equal ($\widehat{y}_{s_6}=\widehat{y}_{s_7}$).

\noindent \textbf{Theorem 1. }Let $t_{\mu}(d)=\mu_{1:d}$ and $r_{\mu}(d)=\gamma_{1:d}$, for $d=1,2,\dots, D$, where $\mu_{i:j}=\sum_{d=i}^{j}\gamma_d\mu_d$, $\gamma_{i:j}=\sum_{d=i}^{j}\gamma_d$ and $t_{\mu}(0)=r_{\mu}(0)=0$. Also, let $G_{\mu}(d)=(r_{\mu}(d), g_{\mu}(d))$ be the GCM points of the cumulative sum diagram with points $(r_{\mu}(d), t_{\mu}(d))$. Define $J_{\mu}^0$ and $J_{\mu}^{1}$ to the indexes of points strictly above $G_{\mu}$ and indexes of its corner points, respectively. Based on the sample $s$, define $t_s(d)=\widehat{y}_{s_{1:d}}$ and $r_s(d)=N_{1:d}$, with $t_{s}(0)=r_{s}(0)=0$, and let $g_s(\cdot)$, $G_s$, $J_s^0$, and $J_s^1$ be the analogous sample quantities of $g_\mu(\cdot)$, $G_\mu$, $J_\mu^0$, and $J_\mu^1$. Denote $A_0$ and $A_1$ to be the events where $J_{\mu}^0 \subseteq J_s^0$ and $J_{\mu}^1 \subseteq J_s^1$, respectively. Then, $P(A_0^c)=o\left(n_N^{-1}\right)$ and $P(A_1^c)=o\left(n_N^{-1}\right)$.\\

To have a better understanding of Theorem 1, note that for every pair of mutually exclusive sets $J_s^0$, $J_s^1$, there are certain poolings (groupings) allowed by $\boldsymbol{\widehat{y}}_s$ to obtain $\boldsymbol{\widehat{\theta}}_s$. In particular, if $J_s^0 \cup J_s^1 = \{1, 2, \dots, D-1\}$ (i.e. no flat spots), then there is a unique pooling allowed by $\boldsymbol{\widehat{y}}_s$. Speaking somewhat loosely and referring to `bad poolings' to those poolings of $\boldsymbol{\widehat{y}}_s$ that are chosen with zero asymptotic probability, Theorem 1 states that bad poolings correspond to those pairs of disjoint sets $J_s^0$, $J_s^1$ that do not satisfy $J_{\mu}^0 \subseteq J_s^0$ and $J_{\mu}^1 \subseteq J_s^1$. 
One case of particular interest is when there are no flat spots on the GCM corresponding to $\boldsymbol{\mu}$, i.e., $J_{\mu}^{0} \cup J_{\mu}^{1}=\{1,2,\dots, D-1\}$. Such scenario is equivalent than saying that, asymptotically, there is a unique pooling allowed by $\boldsymbol{\widehat{y}}_s$. Moreover, under this scenario, it can be proved (Theorem 2) that the proposed CIC$_s$ in Equation \ref{eq:6} is an asymptotic unbiased estimator of the PSE in Equation \ref{eq:7}. \\

\noindent \textbf{Theorem 2. } If $J_{\mu}^{0} \cup J_{\mu}^{1}=\{1,2,\dots, D-1\}$, then $\mathbb{E}[$CIC$_s(\boldsymbol{\widehat{\theta}}_s)]=$PSE$(\boldsymbol{\widehat{\theta}}_s)$+$o(n_N^{-1})$. \\

In practice, the proposed CIC$_s$ can be used as a decision tool that validates the use of the constrained estimator as an estimate of the population domain means. The decision rule would be to choose the estimator, either the constrained or the unconstrained, that produces the smallest CIC$_s$ value. As we mentioned, CIC$_s$ is an overall measure that balances the deviation of the constrained estimator from the unconstrained, as well as the complexity of such estimator. The fact that CIC$_s$ measures the estimator complexity would avoid the undesired situation of choosing always the unconstrained estimator above the constrained estimator.
Although we will focus on the H\'ajek version of the CIC$_s$ (Equation \ref{eq:8}) for the rest of this section, it is important to remark that the following properties are also valid under the Horvitz-Thompson setting. 

Let CIC$_s(\boldsymbol{\tilde{y}}_s)$ and CIC$_s(\boldsymbol{\tilde{\theta}}_s)$ denote the CIC$_s$ values for the unconstrained and constrained estimators, respectively. From Equation \ref{eq:8}, that is,
\begin{align*}
\text{CIC}_s(\boldsymbol{\tilde{y}}_s)&=2 \Tr \left[ \boldsymbol{W}_s \widehat{\text{cov}}(\boldsymbol{\tilde{y}}_s, \boldsymbol{\tilde{y}}_s) \right] , \\
\text{CIC}_s(\boldsymbol{\tilde{\theta}}_s)&=(\boldsymbol{\tilde{y}}_{s}-\boldsymbol{\tilde{\theta}}_s)^{\top} \boldsymbol{W}_s (\boldsymbol{\tilde{y}}_{s}-\boldsymbol{\tilde{\theta}}_s)+2\Tr \left[ \boldsymbol{W}_s \widehat{\text{cov}}(\boldsymbol{\tilde{\theta}}_s, \boldsymbol{\tilde{y}}_s) \right],
\end{align*}
where $\widehat{\text{cov}}(\boldsymbol{\tilde{y}}_{s}, \boldsymbol{\tilde{y}}_{s})_{ij}=\widehat{AC}(\tilde{y}_{s_i}, \tilde{y}_{s_j})$. Similarly as AIC and BIC, we might choose the estimator that produces the smallest CIC$_s$ value. We show that this decision rule is asymptotically correct when choosing the shape based on the limiting domain means $\boldsymbol{\mu}$ (Theorem 5), and also, that the decision made from CIC$_s$ is consistent with the decision made from PSE (Theorem 6). Theorems 3 and 4 contain theoretical properties of $AC(\cdot, \cdot)$ that are required to establish Theorem 5. 

\noindent \textbf{Theorem 3. } \textsl{ For any domains $i_1,i_2,j_1,j_2$ where $i_1\leq j_1$, $i_2 \leq j_2$,}
\begin{equation*}
\underset{N \rightarrow \infty}{\limsup} \;  n_N AC(\tilde{y}_{s_{i_1:j_1}},\tilde{y}_{s_{i_2:j_2}}) < \infty.
\end{equation*}
\textsl{Furthermore,}
\begin{equation*}
n_N\left(\widehat{AC}(\tilde{y}_{s_{i_1:j_1}},\tilde{y}_{s_{i_2:j_2}})-AC(\tilde{y}_{s_{i_1:j_1}},\tilde{y}_{s_{i_2:j_2}})\right) =o_p(1).
\end{equation*}
\\

\noindent \textbf{Theorem 4.} \textsl{ Let $\boldsymbol{\theta}_U=\left( \theta_{U_1}, \theta_{U_2}, \dots, \theta_{U_D}\right)^{\top}$ be the weighted isotonic population domain mean vector of $\overline{\boldsymbol{y}}_U$ with weights $N_1, N_2, \dots, N_D$. Then,} 
\begin{align*}
\tilde{\theta}_{s_d}-\theta_{U_d}&=O_p( n_N^{-1/2}), \; \; \;  \textsl{ for } \; d=1,\dots, D.
\end{align*} 

\noindent \textbf{Theorem 5.}
\begin{center}
	P$\left( \text{CIC}_s(\boldsymbol{\tilde{y}}_s) < \text{CIC}_s(\boldsymbol{\tilde{\theta}}_s) \right) \rightarrow \left\{
	\begin{array}{ll}
	0, & \textsl{ if }  \mu_{1} < \mu_{2} < \dots < \mu_{D}; \\
	1, & \textsl{ if } \mu_1, \mu_2, \dots, \mu_D \textsl{ are not monotone;}
	\end{array} \right. $
\end{center}
\textsl{when} $N\rightarrow \infty$.\\

Theorem 3 states that the scaled $AC(\cdot, \cdot)$ is asymptotically bounded and also, that $\widehat{AC}(\cdot, \cdot)$ is a consistent estimator of $AC(\cdot, \cdot)$ with a rate of $n_N^{-1}$. Hence, both the covariance between $\tilde{y}_{s_{i_1:j_1}}$ and $\tilde{y}_{s_{i_2:j_2}}$, and its proposed estimate are well defined. Theorem 4 establishes that the constrained estimator gets closer to the weighted isotonic population domain mean with a rate of $n_N^{-1/2}$. This theorem generalizes the results in \cite{wu16}, where it was only considered the case when the limiting domain means are monotone. Recall that $\boldsymbol{\theta}_U=\overline{\boldsymbol{y}}_U$ if and only if the population domain means are monotone increasing. Theorem 5 shows that CIC$_s$ consistently chooses the correct estimator based on the order of the limiting domain means $\mu_1, \mu_2, \dots, \mu_D$. 

Finally, Theorem 6 establishes that the chosen estimator driven by PSE in Equation \ref{eq:7} is analogous to the decision made by CIC$_s$. \\

\noindent \textbf{Theorem 6. }
\begin{center}
	$n_N[\text{PSE}(\boldsymbol{\widehat{\theta}}_s) - \text{PSE}(\boldsymbol{\widehat{y}}_s)] \rightarrow \left\{
	\begin{array}{ll}
	0, & \textsl{ if }  \mu_{1} < \mu_{2} < \dots < \mu_{D}; \\
	\infty , & \textsl{ if } \mu_1, \mu_2, \dots, \mu_D \textsl{ are not monotone}.
	\end{array} \right. $
\end{center} 

Observe that neither Theorem 5 nor Theorem 6 deal with the case where the vector entries of $\boldsymbol{\mu}$ are non-strictly monotone. Although in that case we would like both PSE and CIC$_s$ to choose the constrained estimator, neither of them is able to choose it universally. Nevertheless, we show in the Simulations section that the constrained estimator is chosen with a high frequency under the non-strictly monotone scenario.	

\section{Simulations}

We demonstrate the CIC$_s$ performance through simulations under several settings. We consider the set-up in \cite{wu16} as a baseline to produce our simulation scenarios. 
For the first set of simulations, we generate populations of size $N$ using limiting domain means $\mu_1, \dots, \mu_D$. Each element $y_{d_k}$ in the population domain $d$ is independently generated from a normal distribution with mean $\mu_d$ and standard deviation $\sigma$. That is, for a given domain $d$, $y_{d_k} \overset{iid}{\sim} N(\mu_d, \sigma^2)$ for $k=1,2,\dots, N_d$. Samples are generated using a stratified simple random sampling design without replacement in all $H$ strata. The strata constitutes a partition of the total population of size $N$. We make use of an auxiliary random variable $z$ to define the stratum membership of the population elements, with $z$  created by adding random noise $N(0,1)$ to $\sigma (d/D)$, for $d=1,2, \dots, D$. Stratum membership of $y$ is then determined by sorting the vector $z$, creating $H$ blocks of $N/H$ elements based on their ranks, and assigning these blocks to the strata. Also, we set $\sigma=3$, $H=4$, $N_d=N/4$, and $D=4$. The number of replications per simulation is 10000.

The vector of limiting domain means $\boldsymbol{\mu}$ is created using the sigmoid function $S_1(\cdot)$ given by $S_1(d)=2\exp(5d/D-2)/(1+\exp(5d/D-2))$ for $d=1,2,\dots, D$. We consider three different scenarios for $\boldsymbol{\mu}$: the \textsl{monotone scenario}, where $\mu_d$'s are strictly increasing; the \textsl{flat scenario}, where $\mu_d$'s are non-strictly increasing; and the \textsl{non-monotone scenario}, where $\mu_d$'s are not monotone increasing. The limiting domain means on the monotone scenario are given by $\mu_d=S_1(d)$ for $d=1,2,\dots, D$. The flat scenario is formed by ``pulling down'' $\mu_D$ until it is equal to $\mu_{D-1}$, that is,  $\mu_D=S_1(D)-\Delta$ where $\Delta=S_1(D)-S_1(D-1)$. For the non-monotone scenario, we pull $\mu_D$ down until it gets below $\mu_{D-1}$ by using $\mu_D=S_1(D)-2\Delta$. Note that the only difference among these three scenarios relies on the right tail. For each of the above scenarios, the total population size varies from $N=10000, 20000, 40000$. Further, the total sample size $n_N=200N/k$ is divided among the 4 strata as $(25N/k, 50N/k, 50N/k, 100N/k)$ for $k=1000,2000,10000$, which makes the sampling design  informative. Once the sample is generated, the H\'ajek domain mean estimators are computed along with the CIC$_s$ in Equation \ref{eq:8}.

We consider the design Mean Squared Error (MSE) of any estimator $\boldsymbol{\tilde{\phi}}_s$ given by
\begin{equation*}
\text{MSE}(\boldsymbol{\tilde{\phi}}_s)=\mathbb{E}\left[ (\boldsymbol{\tilde{\phi}}_{s}-\boldsymbol{\overline{y}}_U)^{\top} \boldsymbol{W}_U (\boldsymbol{\tilde{\phi}}_{s}-\boldsymbol{\overline{y}}_U) \right].
\end{equation*}
For each scenario mentioned above, we compute both the MSE for the unconstrained estimator MSE$(\boldsymbol{\tilde{y}}_s)$ and for the constrained estimator MSE$(\boldsymbol{\tilde{\theta}}_s)$ through simulations. In addition, we compute the MSE for the CIC$_s$-adaptive estimator $\boldsymbol{\dot{\theta}}_s$, given by
\begin{equation*}
\boldsymbol{\dot{\theta}}_s=\boldsymbol{\tilde{y}}_s I\{\text{CIC}(\boldsymbol{\tilde{y}}_s) < \text{CIC}(\boldsymbol{\tilde{\theta}}_s) \} + \boldsymbol{\tilde{\theta}}_s I\{\text{CIC}(\boldsymbol{\tilde{y}}_s) \geq \text{CIC}(\boldsymbol{\tilde{\theta}}_s) \}.
\end{equation*} 

Although there are no other existing methods that aim to choose between the unconstrained and the constrained estimator for survey data, we compare the performance of CIC$_s$ versus two conditional testing methods that are based on the following hypothesis test under the linear regression model setting,
\begin{equation*}
H_0: \mu_1\leq \mu_2 \leq \dots \leq \mu_D \; \; \; H_1: \text{no restrictions on } \mu_d \;' s.
\end{equation*}
The first test is a naive Wald test which depends on the sample-observed pooling. For this, we compute the test statistic 
\begin{equation*}
Q=(\boldsymbol{\tilde{y}}_s-\boldsymbol{\tilde{\theta}}_{s})^{\top} [\widehat{\text{cov}}(\boldsymbol{\tilde{y}}_s, \boldsymbol{\tilde{y}}_s)]^{-1} (\boldsymbol{\tilde{y}}_s-\boldsymbol{\tilde{\theta}}_{s})
\end{equation*}
and then compare it to a $\chi^2(D-k)$, where $k$ is the number of different estimated values on $\boldsymbol{\tilde{\theta}}_s$.  

The second test is the conditional test proposed by \citet{wollan86}. Even though the latter test is established for independent data with known variances, we use instead the estimated design variances of the sample-observed pooling obtained from Equation \ref{eq:5}. To perform this, we compute the test statistic $Q$ -as in the Wald test- but then we compare it to a $\chi^2(D-k)$ with point mass of $p_0$ at $Q=0$, where $p_0$ is the probability that $Q=0$ under the hypothesis $\mu_1=\mu_2=\dots=\mu_D$. Note that the conditional test might perform similar as the Wald test when the number of domains $D$ is large.

Since both Wald and conditional tests require the variance-covariance matrix of the domain mean estimators to be non-singular, these could be performed only when the variance-covariance matrix formed by the estimates in Equation \ref{eq:5} is in fact a valid covariance matrix. We set the significance level of these tests at $0.05$.

Tables \ref{tab:1}, \ref{tab:2} and \ref{tab:3} contain the proportion of times that the unconstrained estimator is chosen over the constrained estimator under the monotone, flat and non-monotone scenarios, respectively. In cases where the unconstrained and constrained estimators agree (i.e.\ the unconstrained estimator satisfies the constraint), this is counted as a constrained estimator in the calculation of this proportion.  The last two rows of these tables show the MSE of the constrained estimator and the CIC$_s$-adaptive estimator, relative to the MSE of the unconstrained estimator. The former ratio can be viewed as a measure of how much better (or worse) naively applying the constrained estimator is under the different scenarios, while the latter ratio shows how well the adaptive estimator is in terms of balancing the MSE's of the constrained and unconstrained estimators.

From Table \ref{tab:1}, we can note that CIC$_s$ tends not to choose the unconstrained estimator under the monotone scenario as $N$ increases. In contrast, the unconstrained estimator is chosen most of the times under the non-monotone simulation scenario (Table \ref{tab:3}). Flat scenario results (Table \ref{tab:2}) show that although the proportion of times the unconstrained estimator is chosen do not tend to zero as $N$ grows, it is fairly small, meaning that CIC$_s$ is choosing the constrained estimator most of the times. From these three tables, we can observe that CIC$_s$ tends to be more conservative when choosing the unconstrained estimator over the constrained, in comparison with both Wald and conditional tests.

\begin{table}[hp]
	\caption[Simulation results of monotone scenario with $D=4$.]{\textbf{Monotone scenario.} $D=4$. $y_{d_k}$ generated from $N(\mu_d,3^2)$. Based on 10000 replications. Rows 1-3: Proportion of times that unconstrained estimator is chosen using CIC$_s$, Wald test, and conditional test. Rows 4-5: MSE ratios.}
	\label{tab:1}
	\begin{center}
		\resizebox{\textwidth}{!}{
			\begin{tabular}{c|ccc|ccc|ccc}
				$y_{d_k} \sim N(\mu_d,3^2)$ & \multicolumn{3}{c|}{$N=10000$} & \multicolumn{3}{c|}{$N=20000$}  & \multicolumn{3}{c}{$N=40000$} \\ \hline
				& $n=200$ & $n=1000$ & $n=2000$ & $n=400$ & $n=2000$ & $n=4000$ & $n=800$ & $n=4000$ & $n=8000$ \\ \hline
				CIC$_s$  & 0.061 & 0.016 & 0.005 & 0.045 & 0.014 & 0.004 & 0.022 & $4 \times 10^{-4}$ & 0 \\
				Wald   & 0.018 & 0.003 & 0.001 & 0.012 & 0.002 & 0.001 & 0.005 & $10^{-4}$ & 0 \\
				Conditional & 0.020 & 0.004 & 0.001 & 0.013 & 0.003 & 0.001 & 0.005 & $10^{-4}$ & 0 \\ \hline
				MSE$(\boldsymbol{\tilde{\theta}}_s)$/MSE$(\boldsymbol{\tilde{y}}_s)$ & 0.721 & 0.896 & 0.962 & 0.774 & 0.938 & 0.968 & 0.875 & 0.994 & 1 \\
				MSE$(\boldsymbol{\dot{\theta}}_s)$/MSE$(\boldsymbol{\tilde{y}}_s)$ & 0.796 & 0.917 & 0.970 & 0.831 & 0.953 & 0.972 & 0.902 & 0.994 & 1 \\
			\end{tabular}
		}
	\end{center}
\end{table}

\begin{table}[hp]
		\caption[Simulation results of flat scenario with $D=4$.]{\textbf{Flat scenario.} $D=4$. $y_{d_k}$ generated from $N(\mu_d,3^2)$. Based on 10000 replications. Rows 1-3: Proportion of times that unconstrained estimator is chosen using CIC$_s$, Wald test, and conditional test. Rows 4-5: MSE ratios.}
		\label{tab:2}
	\begin{center}
		\resizebox{\textwidth}{!}{
			\begin{tabular}{c|ccc|ccc|ccc}
				$y_{d_k} \sim N(\mu_d,3^2)$ & \multicolumn{3}{c|}{$N=10000$} & \multicolumn{3}{c|}{$N=20000$}  & \multicolumn{3}{c}{$N=40000$} \\ \hline
				& $n=200$ & $n=1000$ & $n=2000$ & $n=400$ & $n=2000$ & $n=4000$ & $n=800$ & $n=4000$ & $n=8000$ \\ \hline
				CIC$_s$    & 0.098 & 0.045  & 0.121 & 0.102 & 0.081 & 0.079 & 0.073 & 0.134 & 0.015 \\
				Wald   & 0.033 & 0.011  & 0.044 & 0.036 & 0.026 & 0.024 & 0.023 & 0.048 & 0.003 \\
				Conditional & 0.038 & 0.013  & 0.047 & 0.040 & 0.029 & 0.026 & 0.025 & 0.052 & 0.004 \\ \hline
				MSE$(\boldsymbol{\tilde{\theta}}_s)$/MSE$(\boldsymbol{\tilde{y}}_s)$ & 0.720 & 0.860 & 0.906 & 0.789 & 0.898 & 0.906 & 0.844 & 0.918 & 0.942\\
				MSE$(\boldsymbol{\dot{\theta}}_s)$/MSE$(\boldsymbol{\tilde{y}}_s)$ & 0.813 & 0.902 & 0.972 & 0.869 & 0.953 & 0.959 & 0.901 & 0.985 & 0.959 \\
			\end{tabular}
		}
	\end{center}
\end{table}

\begin{table}[hp]
	\caption[Simulation results of non-monotone scenario with $D=4$.]{\textbf{Non-monotone scenario.} $D=4$. $y_{d_k}$ generated from $N(\mu_d,3^2)$. Based on 10000 replications. Rows 1-3: Proportion of times that unconstrained estimator is chosen using CIC$_s$, Wald test, and conditional test. Rows 4-5: MSE ratios.}
	\label{tab:3}
	\begin{center}
		\resizebox{\textwidth}{!}{
			\begin{tabular}{c|ccc|ccc|ccc}
				$y_{d_k} \sim N(\mu_d,3^2)$ & \multicolumn{3}{c|}{$N=10000$} & \multicolumn{3}{c|}{$N=20000$}  & \multicolumn{3}{c}{$N=40000$} \\ \hline
				& $n=200$ & $n=1000$ & $n=2000$ & $n=400$ & $n=2000$ & $n=4000$ & $n=800$ & $n=4000$ & $n=8000$ \\ \hline
				CIC$_s$    & 0.118 & 0.126  & 0.602 & 0.126 & 0.497 & 0.513 & 0.172 & 0.623 & 0.963 \\
				Wald   & 0.042 & 0.045  & 0.386 & 0.051 & 0.299 & 0.302 & 0.070 & 0.420 & 0.894 \\
				Conditional & 0.048 & 0.049  & 0.403 & 0.056 & 0.310 & 0.315 & 0.073 & 0.434 & 0.899 \\ \hline
				MSE$(\boldsymbol{\tilde{\theta}}_s)$/MSE$(\boldsymbol{\tilde{y}}_s)$ & 0.712 & 0.854 & 1.346 & 0.695 & 1.211 & 1.224 & 0.860 & 1.400 & 2.705 \\
				MSE$(\boldsymbol{\dot{\theta}}_s)$/MSE$(\boldsymbol{\tilde{y}}_s)$ & 0.814 & 0.928 & 1.128 & 0.807 & 1.115 & 1.118 & 0.945 & 1.137 & 1.037 \\
			\end{tabular}
		}
	\end{center}
\end{table}

On a second set of simulations, we consider the case where the population elements are generated from a skewed distribution. For a given domain $d$, $y_{d_k}$ is generated from a $\chi^2$ distribution with $\mu_d$ degrees of freedom, for $k=1,2,\dots, N_d$ and $D=4$. As in the first set of simulations, we consider the same three scenarios for $\boldsymbol{\mu}$ using the $S_1(\cdot)$ sigmoid function. For each of them, we consider the case where $N=10000$ and $n_N=200, 1000, 2000$. Table \ref{tab:4} contains the results of this skewed case. Again, we can observe that CIC$_s$ behaves as expected despite the skewness of the population generating distribution.

\begin{table}[hp]
	\caption[Simulation results of the skewed case with $D=4$.]{\textbf{Skewed case.} $D=4$. $y_{d_k}$ generated from $\chi^2(\mu_d)$. Based on 10000 replications. Rows 1-3: Proportion of times that unconstrained estimator is chosen using CIC$_s$, Wald test, and conditional test. Rows 4-6: MSE ratios.}
	\label{tab:4}
	\begin{center}
		\resizebox{\textwidth}{!}{
			\begin{tabular}{c|ccc|ccc|ccc}
				$y_{d_k} \sim \chi^2(\mu_d)$ & \multicolumn{3}{c|}{Monotone} & \multicolumn{3}{c|}{Flat}  & \multicolumn{3}{c}{Non-monotone} \\ \hline
				& $n=200$ & $n=1000$ & $n=2000$ & $n=200$ & $n=1000$ & $n=2000$ & $n=200$ & $n=1000$ & $n=2000$ \\ \hline
				CIC$_s$    	& 0.029 & 0.003  & 0 & 0.052 & 0.079 & 0.138 & 0.193 & 0.326 & 0.693 \\
				Wald   		& 0.014 & 0.001  & 0 & 0.024 & 0.031 & 0.055 & 0.114 & 0.172 & 0.573 \\
				Conditional & 0.014 & 0.001  & 0 & 0.025 & 0.033 & 0.057 & 0.117 & 0.177 & 0.579 \\ \hline
				MSE$(\boldsymbol{\tilde{\theta}}_s)$/MSE$(\boldsymbol{\tilde{y}}_s)$ & 0.808 & 0.958 & 1 & 0.806 & 0.853 & 0.886 & 0.817 & 1.034 & 1.890 \\
				MSE$(\boldsymbol{\dot{\theta}}_s)$/MSE$(\boldsymbol{\tilde{y}}_s)$ & 0.855 & 0.966 &  1 & 0.872 & 0.936 & 0.982 & 0.927 & 1.086 & 1.230 \\
			\end{tabular}
		}
	\end{center}
\end{table}

A third set of simulations considers the case where the domain mean estimators are more correlated in comparison with the stratified simple random sample simulations. The setting for this simulation set is basically equal to the first set, except that we use the auxiliary variable $z$ to create 100 clusters. Then, we sample $r$ clusters with equal probability. We let $r=2,10,20$. We only consider the case where $N=10000$ and $n_N=200,1000,2000$ for each of the three scenarios. Table \ref{tab:5} contains the simulation results for this correlated case. Note that CIC$_s$ is choosing the unconstrained estimator with a low proportion under the monotone scenario, which is desired. However, the proportion of times that the unconstrained estimator is chosen under the non-monotone scenario is almost half in comparison of its corresponding stratified simple random sample simulation (see Table \ref{tab:3}). The stars ($*$) in Table \ref{tab:5} mean that results for the Wald and the conditional tests are not available since the estimated variance-covariance matrix of the H\'ajek domain means is in fact a singular matrix. Recall that both tests need such matrix to be a valid covariance matrix in order to be performed. Note that on those cases with stars, CIC$_s$ continues to be a plausible option to choose between the two estimators.

\begin{table}[hp]
	\caption[Simulation results of the correlated case with $D=4$.]{\textbf{Correlated case.} $D=4$. $y_{d_k}$ generated from $N(\mu_d,3^3)$. Based on 10000 replications. Rows 1-3: Proportion of times that unconstrained estimator is chosen using CIC$_s$, Wald test, and conditional test. Rows 4-6: MSE ratios.}
	\label{tab:5}
	\begin{center}
		\resizebox{0.9\textwidth}{!}{
			\begin{tabular}{c|ccc|ccc|ccc}
				$y_{d_k} \sim N(\mu_d,3^2)$ & \multicolumn{3}{c|}{Monotone} & \multicolumn{3}{c|}{Flat}  & \multicolumn{3}{c}{Non-monotone} \\ \hline
				& $r=2$ & $r=10$ & $r=20$ & $r=2$ & $r=10$ & $r=20$ & $r=2$ & $r=10$ & $r=20$ \\ \hline
				CIC$_s$ 	 & 0.194 & 0.025 & 0.005 & 0.245 & 0.085 & 0.069 & 0.284 & 0.461 & 0.696 \\
				Wald    	 &   * 	& 0.011  & 0.001 &   * 	 & 0.071 & 0.035 &   *	 & 0.417 & 0.574 \\
				Conditional  &   * 	& 0.019  & 0.002 &   * 	 & 0.072 & 0.037 &   *	 & 0.422 & 0.582 \\ \hline
				MSE$(\boldsymbol{\tilde{\theta}}_s)$/MSE$(\boldsymbol{\tilde{y}}_s)$ & 0.717 & 0.901 & 0.958 & 0.690 & 0.838 & 0.842 & 0.694 & 1.263 & 1.911 \\
				MSE$(\boldsymbol{\dot{\theta}}_s)$/MSE$(\boldsymbol{\tilde{y}}_s)$ & 0.862 & 0.937 &  0.966 & 0.836 & 0.930 & 0.929 & 0.856 & 1.178 & 1.233 \\
			\end{tabular}
		}
	\end{center}
\end{table} 

Table \ref{tab:5} shows that although the CIC$_s$ performs as expected for the correlated case, the unconstrained estimator is being chosen only 69.6\% of the times under the non-monotone scenario when the sample size is 20\% of the total population. One plausible reason could be the fact that the monotonicity violation on this scenario is weak. Therefore, we would like to analyze the efficacy of the CIC$_s$ as the violation of monotonicity increases. To do that, we consider again the correlated case. To increase the violation on the limiting domain means, we create $\mu_D$ from pulling down $S_1(D)$ by a quantity $t\Delta$, where $t=3,4,5$. That is, $\mu_D=S_1(D)-t\Delta$. The results of this simulation case (Table \ref{tab:6}) shows that the MSE ratio between the unconstrained and the constrained estimators overpass 1 as the violation increases. Moreover, the proportion of times that the unconstrained estimator is chosen also increases and approaches to 1 as expected.

\begin{table}[hp]
	\caption[Simulation results of the correlated case when the monotonicity violation is increased, with $D=4$.]{\textbf{Increasing Monotonicity Violation - Correlated case.} $D=4$. $y_{d_k}$ generated from $N(\mu_d,3^2)$. Based on 10000 replications. Rows 1-3: Proportion of times that unconstrained estimator is chosen using CIC$_s$, Wald test, and conditional test. Rows 4-6: MSE ratios.}
	\label{tab:6}
	\begin{center}
		\resizebox{0.9\textwidth}{!}{
			\begin{tabular}{c|ccc|ccc|ccc}
				$y_{d_k} \sim N(\mu_d,3^2)$ & \multicolumn{3}{c|}{$\mu_D=S_1(D)-3\Delta$} & \multicolumn{3}{c|}{$\mu_D=S_1(D)-4\Delta$}  & \multicolumn{3}{c}{$\mu_D=S_1(D)-5\Delta$} \\ \hline
				& $r=2$ & $r=10$ & $r=20$ & $r=2$ & $r=10$ & $r=20$ & $r=2$ & $r=10$ & $r=20$ \\ \hline
				CIC$_s$ 	& 0.388 & 0.708  & 0.934 & 0.450 & 0.881 & 0.936 &  0.507 & 0.963 & 1 \\
				Wald   		&   *   & 0.658  & 0.882 &   *   & 0.852 & 0.835 &    *   & 0.952 & 1 \\
				Conditional &   *   & 0.664  & 0.885 &   *   & 0.854 & 0.890 &    *   & 0.953 & 1 \\ \hline
				MSE$(\boldsymbol{\tilde{\theta}}_s)$/MSE$(\boldsymbol{\tilde{y}}_s)$ & 0.798 & 1.963 & 3.554 & 0.882 & 2.999 & 3.617 & 1.022 & 4.302 & 9.037 \\
				MSE$(\boldsymbol{\dot{\theta}}_s)$/MSE$(\boldsymbol{\tilde{y}}_s)$ & 0.962 & 1.233 & 1.107 & 1.002 & 1.169 & 1.109 & 1.059 & 1.081 & 1.000 \\
			\end{tabular}
		}
	\end{center}
\end{table}

We also perform simulations to study the behavior of CIC$_s$ when the number of domains is larger than 4. We consider the case where $D=8$. The values of $\boldsymbol{\mu}$ are obtained from the sigmoid function $S_2(d)=4\exp(5d/D-2)/(1+\exp(5d/D-2))$. The setting in this 8-domain case is basically the same as the first simulation set, but using $S_2(\cdot)$ instead of $S_1(\cdot)$, $N=20000$, and $n_N=400, 2000, 40000$. We choose these values for $N$ and $n_N$ in order to have a similar rough average sample size in each domain as it was in simulations where $D=4$. As shown in Table \ref{tab:7}, CIC$_s$ follows a similar behavior as in the previous simulations. 

\begin{table}[hp]
	\caption[Simulation results when $D=8$.]{\textbf{8-domain case.} $D=8$. $y_{d_k}$ generated from $N(\mu_d,3^2)$. Based on 10000 replications. Rows 1-3: Proportion of times that unconstrained estimator is chosen using CIC$_s$, Wald test, and conditional test. Rows 4-5: MSE ratios.}
	\label{tab:7}
	\begin{center}
		\resizebox{\textwidth}{!}{
			\begin{tabular}{c|ccc|ccc|ccc}
				$y_{d_k} \sim N(\mu_d,3^2)$ & \multicolumn{3}{c|}{Monotone} & \multicolumn{3}{c|}{Flat}  & \multicolumn{3}{c}{Non-monotone} \\ \hline
				& $n=400$ & $n=2000$ & $n=4000$ & $n=400$ & $n=2000$ & $n=4000$ & $n=400$ & $n=2000$ & $n=4000$ \\ \hline
				CIC$_s$  & 0.054 & 0.042  & 0.003 & 0.075 & 0.127 & 0.060 & 0.084 & 0.287 & 0.631 \\
				Wald   	 & 0.021 & 0.010  & $4 \times 10^{-4}$ & 0.031 & 0.048 & 0.017 & 0.037 & 0.158 & 0.439 \\
				Conditional & 0.023 & 0.010  & $4 \times 10^{-4}$ & 0.034 & 0.049 & 0.017 & 0.041 & 0.159 & 0.441 \\ \hline
				MSE$(\boldsymbol{\tilde{\theta}}_s)$/MSE$(\boldsymbol{\tilde{y}}_s)$ & 0.666 & 0.902 & 0.975 & 0.648 & 0.877 & 0.961 & 0.666 & 0.935 & 1.162 \\
				MSE$(\boldsymbol{\dot{\theta}}_s)$/MSE$(\boldsymbol{\tilde{y}}_s)$ & 0.719 & 0.921 &  0.978 & 0.710 & 0.918 & 0.978 & 0.731 & 0.970 & 1.047 \\
			\end{tabular}
		}
	\end{center}
\end{table}

We end this section by showing simulation results obtained using the exact same set-up as in \cite{wu16}. To get the $\mu_d$ values, we use the sigmoid function $S_3(d)=\exp(20d/D-10)/(1+\exp(20d/D-10))$. We set the population size as $N=1000$ and the domain size as $N_d=N/D$. We simulate the $y_{d_k}$ values from a normal distribution with mean $\mu_d$ and standard deviation $\sigma$. As it was done before, samples are generated from a stratified sampling design with simple random sampling without replacement in each of four strata; and the stratum membership was assigned using the auxiliary random variable $z$. 

We study four cases obtained by varying the number of domains $D=5,20$; and the standard deviation $\sigma=0.5,1$. The sample size is set to $n_N=200$ when $D=5$, splitted as 25, 50, 50, 75 samples in each stratum; and $n_N=800$ when $D=20$, splitted as 100, 200, 200, 300 samples in each stratum. For each case, we create 7 different cases for $\boldsymbol{\mu}$. These cases are determined by setting $\mu_d=S_3(d)$ for $d=1,\dots, D-1$; and $\mu_D=S_3(D-1)-\delta$ for $\delta=0, \pm 0.15, \pm 0.3, \pm 0.45$. Note that $\delta=0$ corresponds to the flat scenario, meanwhile $\delta<0$ define monotone scenarios and $\delta>0$ define non-monotone scenarios.

\begin{figure}[htp]
	\subfigure[$\sigma=0.5$, $D=5$.]{\includegraphics[width=.5\textwidth]{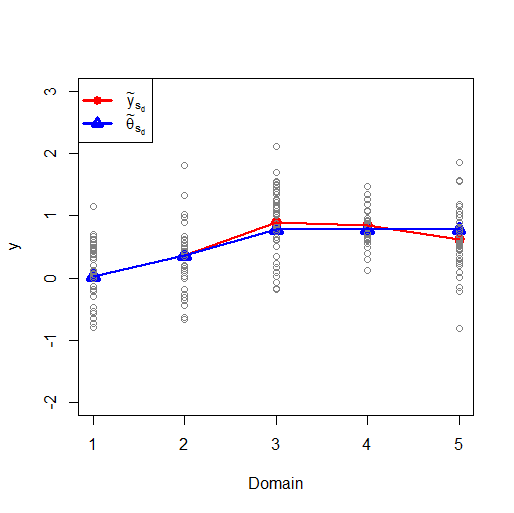}}\
	\subfigure[$\sigma=0.5$, $D=20$.]{\includegraphics[width=.5\textwidth]{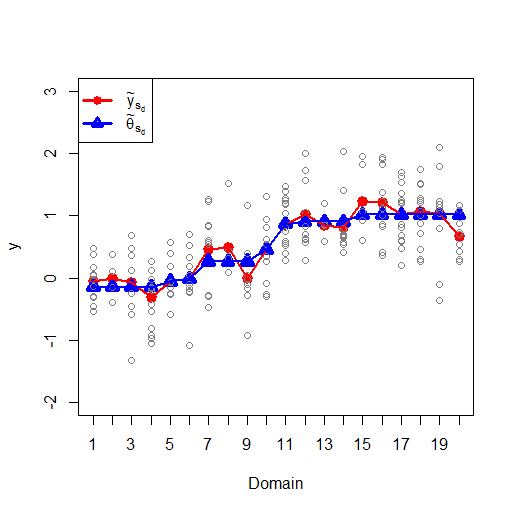}}\\
	\subfigure[$\sigma=1$, $D=5$.]{\includegraphics[width=.5\textwidth]{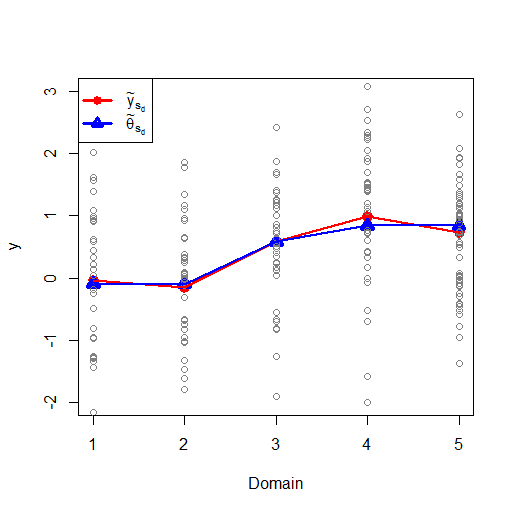}}\
	\subfigure[$\sigma=1$, $D=20$.]{\includegraphics[width=.5\textwidth]{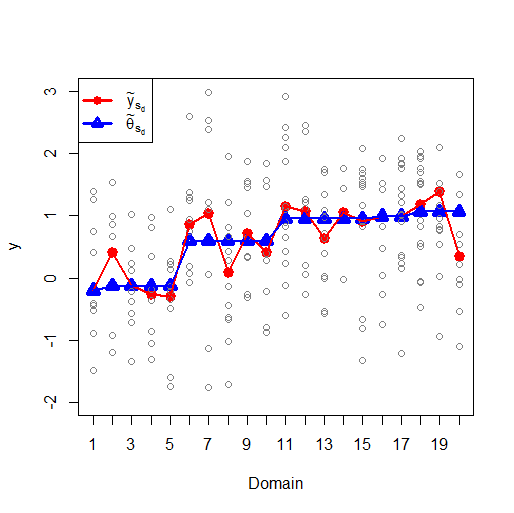}}\\
	\caption{One fitted samples for each of four cases obtained using $S_3(\cdot)$. Dots correspond to unconstrained estimates, triangles to constrained estimates. }
	\label{fig:2}
\end{figure}

\begin{figure}[htp]
	\subfigure[$D=5$.]{\label{fig:3.1} \includegraphics[width=.5\textwidth]{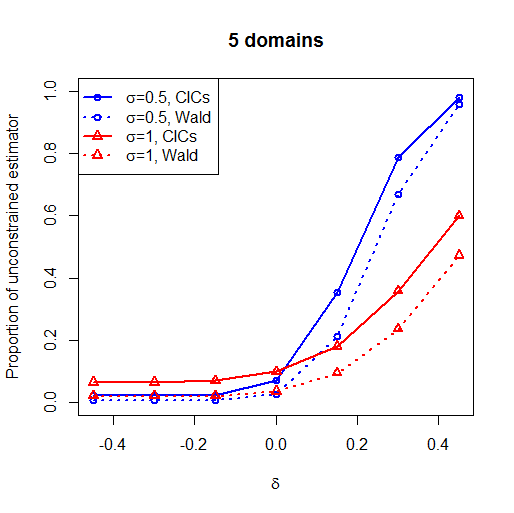}}\
	\subfigure[$D=20$.]{\label{fig:3.2} \includegraphics[width=.5\textwidth]{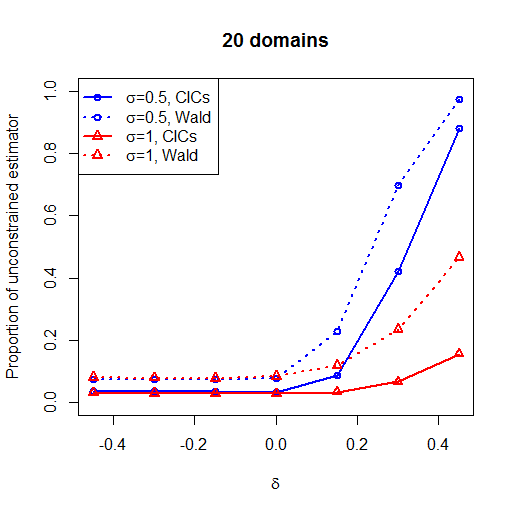}}\\
	\caption{Proportion of times that the unconstrained estimator is chosen under the 4 scenarios of $S_3(\cdot)$, for several values of $\delta$. Solid lines: CIC$_s$, dotted lines: Wald test. Dots: $\sigma=0.5$, triangles: $\sigma=1$.}
	\label{fig:3}
\end{figure}

Figure \ref{fig:2} contains examples of one fitted samples for each of the four cases mentioned above. Note that the fact that the $S_3$ sigmoid function is considerably flat at its extremes makes especially complicated to decide whether the population domain means are isotonic or not, when $D=20$. Tables \ref{tab:8}-\ref{tab:11} present the proportion of times that the unconstrained estimator is chosen in each case, along with MSE ratios. To visualize these results better, we create Figure \ref{fig:3} which contains plots of the proportion of times that the unconstrained estimator is chosen under CIC$_s$ and Wald test, for the set values of $\delta$. We ignore the results obtained by the conditional test since these are shown to be practically the same as lead by the Wald test (see Tables \ref{tab:8}-\ref{tab:11}).

Plots in Figure \ref{fig:3} demonstrates that both CIC$_s$ and Wald test perform better when the standard deviation is smaller. Figure \ref{fig:3.1} shows that CIC$_s$ tends to choose more the unconstrained estimator than the Wald test, when $D=5$. This fact provides evidence that the CIC$_s$ does better than the Wald test under non-monotone scenarios. In contrast, Figure \ref{fig:3.2} shows an opposite behavior between CIC$_s$ and Wald test. The worst performance for both CIC$_s$ and Wald test is shown when $D=20$ and $\sigma=1$. In this case, CIC$_s$ chooses the constrained estimator more than 80\% of times, meanwhile Wald test choose it a little less than 60\% of times, although is desirable to never choose it. However, it can be seen in Table 11 that the MSE ratio of the constrained estimator over the unconstrained estimator does not show neither a clear preference for the latter estimator. 

\begin{table}[hp]
	\caption[Simulation results when $D=5$ and $\sigma=0.5$.]{\boldsymbol{$S_3(\cdot), \; D=5, \; \sigma=0.5$}. $n_N=200$. Based on 10000 replications. Rows 1-3: Proportion of times that unconstrained estimator is chosen using CIC$_s$, Wald test, and conditional test. Rows 4-5: MSE ratios.}
	\label{tab:8}
	\begin{center}
		\resizebox{\textwidth}{!}{
			\begin{tabular}{c|ccc|c|ccc}
				$y_{d_k} \sim N(\mu_d,0.5^2)$ & \multicolumn{3}{c|}{Monotone} & \multicolumn{1}{c|}{Flat}  & \multicolumn{3}{c}{Non-monotone} \\ \hline
				& $\delta=-0.45$ & $\delta=-0.30$ & $\delta=-0.15$ & $\delta=0$ & $\delta=0.15$ & $\delta=0.30$ & $\delta=0.45$ \\ \hline
				CIC$_s$  & 0.023 & 0.023  & 0.024 & 0.072 & 0.352 & 0.787 & 0.980 \\
				Wald   	 & 0.006 & 0.006  & 0.006 & 0.026 & 0.212 & 0.667 & 0.958 \\
				Conditional & 0.006 & 0.006  & 0.006 & 0.026 & 0.213 & 0.668 & 0.959 \\ \hline
				MSE$(\boldsymbol{\tilde{\theta}}_s)$/MSE$(\boldsymbol{\tilde{y}}_s)$ & 0.882 & 0.880 & 0.857 & 0.781 & 0.957 & 1.822 & 3.479 \\
				MSE$(\boldsymbol{\dot{\theta}}_s)$/MSE$(\boldsymbol{\tilde{y}}_s)$ & 0.911 & 0.909 &  0.887 & 0.849 & 1.013 & 1.153 & 1.036 \\
			\end{tabular}
		}
	\end{center}
\end{table}

\begin{table}[hp]
	\caption[Simulation results when $D=5$ and $\sigma=1$.]{\boldsymbol{$S_3(\cdot), \; D=5, \; \sigma=1$}. $n_N=200$. Based on 10000 replications. Rows 1-3: Proportion of times that unconstrained estimator is chosen using CIC$_s$, Wald test, and conditional test. Rows 4-5: MSE ratios.}
	\label{tab:9}
	\begin{center}
		\resizebox{\textwidth}{!}{
			\begin{tabular}{c|ccc|c|ccc}
				$y_{d_k} \sim N(\mu_d,1^2)$ & \multicolumn{3}{c|}{Monotone} & \multicolumn{1}{c|}{Flat}  & \multicolumn{3}{c}{Non-monotone} \\ \hline
				& $\delta=-0.45$ & $\delta=-0.30$ & $\delta=-0.15$ & $\delta=0$ & $\delta=0.15$ & $\delta=0.30$ & $\delta=0.45$ \\ \hline
				CIC$_s$  & 0.065 & 0.065  & 0.070 & 0.099 & 0.181 & 0.358 & 0.600 \\
				Wald   	 & 0.021 & 0.021  & 0.021 & 0.036 & 0.095 & 0.236 & 0.473 \\
				Conditional & 0.022 & 0.021  & 0.021 & 0.036 & 0.095 & 0.237 & 0.474 \\ \hline
				MSE$(\boldsymbol{\tilde{\theta}}_s)$/MSE$(\boldsymbol{\tilde{y}}_s)$ & 0.806 & 0.788 & 0.747 & 0.704 & 0.732 & 0.915 & 1.296 \\
				MSE$(\boldsymbol{\dot{\theta}}_s)$/MSE$(\boldsymbol{\tilde{y}}_s)$ & 0.875 & 0.858 &  0.826 & 0.807 & 0.861 & 1.012 & 1.145 \\
			\end{tabular}
		}
	\end{center}
\end{table}

\begin{table}[hp]
	\caption[Simulation results when $D=20$ and $\sigma=0.5$.]{\boldsymbol{$S_3(\cdot), \; D=20, \; \sigma=0.5$}. $n_N=800$. Based on 10000 replications. Rows 1-3: Proportion of times that unconstrained estimator is chosen using CIC$_s$, Wald test, and conditional test. Rows 4-5: MSE ratios.}
	\label{tab:10}
	\begin{center}
		\resizebox{\textwidth}{!}{
			\begin{tabular}{c|ccc|c|ccc}
				$y_{d_k} \sim N(\mu_d,0.5^2)$ & \multicolumn{3}{c|}{Monotone} & \multicolumn{1}{c|}{Flat}  & \multicolumn{3}{c}{Non-monotone} \\ \hline
				& $\delta=-0.45$ & $\delta=-0.30$ & $\delta=-0.15$ & $\delta=0$ & $\delta=0.15$ & $\delta=0.30$ & $\delta=0.45$ \\ \hline
				CIC$_s$  & 0.037 & 0.037  & 0.036 & 0.034 & 0.087 & 0.422 & 0.881 \\
				Wald   	 & 0.074 & 0.074  & 0.073 & 0.078 & 0.229 & 0.697 & 0.972 \\
				Conditional & 0.074 & 0.074  & 0.073 & 0.078 & 0.229 & 0.697 & 0.972 \\ \hline
				MSE$(\boldsymbol{\tilde{\theta}}_s)$/MSE$(\boldsymbol{\tilde{y}}_s)$ & 0.503 & 0.503 & 0.495 & 0.468 & 0.556 & 0.905 & 1.533 \\
				MSE$(\boldsymbol{\dot{\theta}}_s)$/MSE$(\boldsymbol{\tilde{y}}_s)$ & 0.539 & 0.539 &  0.530 & 0.503 & 0.625 & 0.994 & 1.075 \\
			\end{tabular}
		}
	\end{center}
\end{table}

\begin{table}[hp]
	\caption[Simulation results when $D=20$ and $\sigma=1$.]{\boldsymbol{$S_3(\cdot), \; D=20, \; \sigma=1$}. $n_N=800$. Based on 10000 replications. Rows 1-3: Proportion of times that unconstrained estimator is chosen using CIC$_s$, Wald test, and conditional test. Rows 4-5: MSE ratios.}
	\label{tab:11}
	\begin{center}
		\resizebox{\textwidth}{!}{
			\begin{tabular}{c|ccc|c|ccc}
				$y_{d_k} \sim N(\mu_d,1^2)$ & \multicolumn{3}{c|}{Monotone} & \multicolumn{1}{c|}{Flat}  & \multicolumn{3}{c}{Non-monotone} \\ \hline
				& $\delta=-0.45$ & $\delta=-0.30$ & $\delta=-0.15$ & $\delta=0$ & $\delta=0.15$ & $\delta=0.30$ & $\delta=0.45$ \\ \hline
				CIC$_s$  & 0.031 & 0.030  & 0.028 & 0.028 & 0.034 & 0.067 & 0.156 \\
				Wald   	 & 0.081 & 0.079  & 0.078 & 0.084 & 0.119 & 0.235 & 0.466 \\
				Conditional & 0.081 & 0.079  & 0.078 & 0.084 & 0.119 & 0.235 & 0.466 \\ \hline
				MSE$(\boldsymbol{\tilde{\theta}}_s)$/MSE$(\boldsymbol{\tilde{y}}_s)$ & 0.415 & 0.410 & 0.398 & 0.386 & 0.402 & 0.475 & 0.617 \\
				MSE$(\boldsymbol{\dot{\theta}}_s)$/MSE$(\boldsymbol{\tilde{y}}_s)$ & 0.451 & 0.445 &  0.431 & 0.420 & 0.441 & 0.540 & 0.723 \\
			\end{tabular}
		}
	\end{center}
\end{table}

\newpage

\section{Real data application: NHANES data}
We apply the proposed CIC$_s$ methodology to the 2011-2012 NHANES laboratory data obtained from the Center of Disease Control website. There are $n_N=1637$ complete observations for variables age and LDL-cholesterol measures (mg/dL), where we only consider observations with age range between 21-60 years old. The LDL-cholesterol measure is the variable of interest $y$. Under the consideration that LDL-cholesterol measures might increase with age, we intend to use that information on the construction of domain means estimates. We create 10 domains by partitioning the age variable in 10 categories of three years each, i.e., 21-24, 25-28, \dots, 57-60. 

Since there is no information available regard the population domain sizes $N_d$, we compute both unconstrained and constrained estimators of the population domain means using the H\'ajek estimator. The constrained estimator in Equation \ref{eq:3} is obtained by using the PAVA. The covariance term in CIC$_s$ for both estimators is estimated using Equation \ref{eq:5}.

\begin{figure}[htp]
	\begin{center}
		\includegraphics[width=0.70\textwidth]{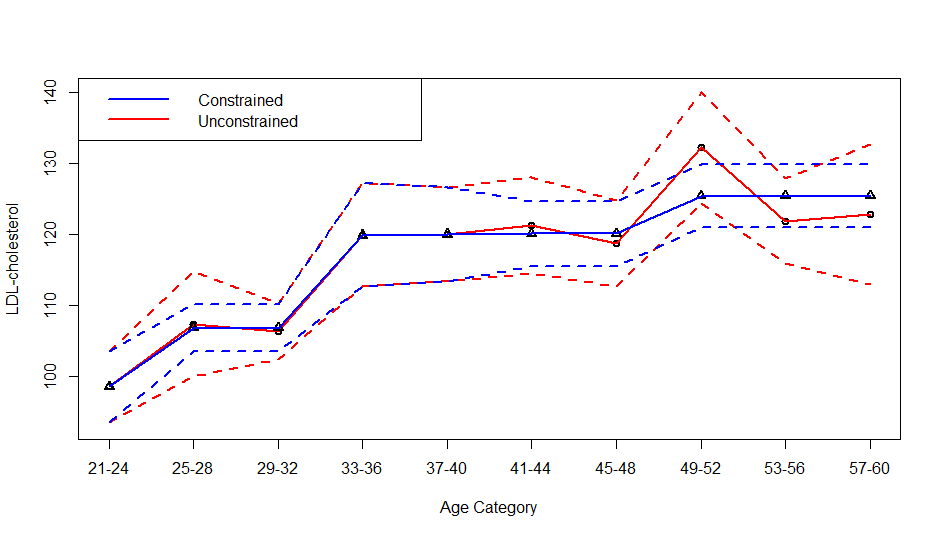}
	\end{center}
	\caption{2011-2012 NHANES laboratory data. Solid lines: constrained and unconstrained estimators. Dotted lines: pointwise 95\% Wald confidence intervals. $\text{CIC}_s(\boldsymbol{\tilde{y}}_s)=23.354$, $\text{CIC}_s(\boldsymbol{\tilde{\theta}}_s)=18.874$.}
	\label{fig:4}
\end{figure}

Figure \ref{fig:4} contains both unconstrained and constrained estimators along with their pointwise 95\% Wald confidence intervals. The variance estimates to construct these intervals are based on Equation \ref{eq:5}, and the observed pooling is used to compute the estimated variance of the constrained estimator. Note that there are notable differences between them on the last three domains. Since $\text{CIC}_s(\boldsymbol{\tilde{y}}_s)=23.354$ and $\text{CIC}_s(\boldsymbol{\tilde{\theta}}_s)=18.874$, then our proposed method chooses the constrained estimator above the unconstrained as an estimate of the population domain means. Moreover, notice that the confidence interval is tighter for the constrained estimator than for the unconstrained, which shows the fact that pooling domains decrease the uncertainty of the estimates.

\newpage

\section{Conclusions}
We have proposed the Cone Information Criterion for Survey data (CIC$_s$) as a data-driven criterion for choosing between the constrained and the unconstrained domain mean estimators. We showed that the CIC$_s$ is consistently selecting the correct estimator based on the shape of the limiting domain means $\boldsymbol{\mu}$. Moreover, the CIC$_s$ shares similar characteristics with other information criteria like AIC and BIC. Mainly, it is a measure that balances the deviation of the constrained estimator from the unconstrained with a measure of the complexity of such estimator. 

Some generalizations can be naturally derived from this work. Note the trace term in the CIC$_s$ could be multiplied by any positive constant $C$ (instead of $C=2$, as proposed) so that the consistency of the CIC$_s$ remains true. The larger the value of $C$ would imply a larger penalization of the constrained estimator complexity. Since we are able to control the amount of such penalization by changing the value of $C$, one question might be how to choose the optimum value $C$. A generalization of a more practical interest might be to extend the CIC$_s$ to other shape constraints beyond monotonicity, so that it can be used to choose among many other types of shapes on the survey context. In that case, the constrained estimator might be computed through the Cone Projection Algorithm proposed by \cite{meyer13b}. Both of these extensions are currently being considered by the authors.

\section*{Acknowledgment} 

This material is based upon research work supported by the National Science Foundation under Grant MMS-1533804.

\bibliography{CristianOliva_bib}{}
\bibliographystyle{apalike}

\section*{Appendix}

The first part of this section contains all lemmas used to prove the theoretical results contained in Sections 2 and 3. Complete proofs of latter results are included at the end of this section. \\
	
	\textbf{Lemma 1.}	
	$\mathbb{E}[(\widehat{y}_{s_{i:j}}-\overline{y}_{U_{i:j}})^4]=o(n_N^{-1})$, for any $i\leq j$, $i,j=1,2,\dots,D$.
	\begin{proof}[Proof of Lemma 1]
		For simplicity of notation and without loss of generality, we will use $s$ instead of $s_{i:j}$ and $U$ instead of $U_{i:j}$. Note that
		\begin{align*}
		n_N\mathbb{E}[(\widehat{y}_{s}-\overline{y}_{U})^4] &= \frac{n_N}{N^4}\sum_{k \in U} \sum_{l \in U} \frac{y_k^2y_l^2}{\pi_k^2\pi_l^2} \mathbb{E}[(I_k-\pi_k)^2(I_l-\pi_l)^2] \\
		&+ \frac{n_N}{N^4}\sum_{k \in U} \sum_{p,q \in U:p \neq q} \frac{y_k^2 y_p y_q}{\pi_k^2\pi_p\pi_q} \mathbb{E}[(I_k-\pi_k)^2(I_p-\pi_p)(I_q-\pi_q)] \\
		&+ \frac{n_N}{N^4}\sum_{k,l \in U:k\neq l} \sum_{p,q \in U:p \neq q} \frac{y_k y_l y_p y_q}{\pi_k\pi_l\pi_p\pi_q} \mathbb{E}[(I_k-\pi_k)(I_l-\pi_l)(I_p-\pi_p)(I_q-\pi_q)] \\
		&=c_{1N}+c_{2N}+c_{3N}.
		\end{align*}
		We will now prove that $c_{1N}, c_{2N}, c_{3N}$ converge to zero as $N$ goes to infinity. For $c_{1N}$, we have that
		\begin{align*}
		|c_{1N}|&\leq \frac{n_N}{N^4} \sum_{k \in U} \frac{y_k^4}{\pi_k^4}\mathbb{E}\left[(I_k-\pi_k)^4 \right] +\frac{n_N}{N^4} \sum_{(k,l) \in D_{2,N}} \frac{y_k^2y_l^2}{\pi_k^2\pi_l^2}\mathbb{E}\left[(I_k-\pi_k)^2(I_l-\pi_l)^2 \right] \\ &\leq \frac{n_N}{N \lambda^4} \frac{\sum_{k \in U} y_k^4}{N} \left(\frac{1}{N^2}+\frac{1}{N}\right),
		\end{align*} 
		where the term to the right goes to zero from Assumptions (A2)-(A3). Further, 
		\begin{align*}
		|c_{2N}|&\leq \frac{2 n_N}{N^4} \sum_{(k,p) \in D_{2,N}} \frac{y_k^3y_p}{\pi_k^3 \pi_p} |\mathbb{E}[(I_k-\pi_k)^3(I_p-\pi_p)] | \\
		&+ \frac{n_N}{N^4} \sum_{(k,p,q) \in D_{3,N}} \frac{y_k^2y_py_q}{\pi_k \pi_p \pi_q} |\mathbb{E}[(I_{k}-\pi_{k})^2(I_{p}-\pi_{p})(I_{q}-\pi_{q}) ]|  \\
		&\leq \frac{n_N}{N\lambda^4}\frac{\sum_{k \in U} y_k^4 }{N} \left( \frac{2}{N} +  \underset{(k,p,q) \in D_{3,N}}{\max}|\mathbb{E}[(I_{k}-\pi_{k})^2(I_{p}-\pi_{p})(I_{q}-\pi_{q}) ]| \right), 
		\end{align*}
		which converges to zero by Assumption (A7). Finally, note that
		\begin{align*}
		|c_{3N}| &\leq \frac{2 n_N}{N^4} \sum_{(k,l) \in D_{2,N} } \frac{y_k^2y_l^2}{\pi_k^2\pi_l^2} |\mathbb{E}[(I_{k}-\pi_{k})^2(I_{l}-\pi_{l})^2 ]| \\
		&+ \frac{2 n_N}{N^4}  \sum_{(k,l,p) \in D_{3,N}}  \frac{y_k^2y_ly_q}{\pi_k^2\pi_l\pi_q} |\mathbb{E}[(I_{k}-\pi_{k})^2(I_{l}-\pi_{l})(I_{q}-\pi_{q}) ]| \\
		&+ \frac{n_N}{N^4} \sum_{(k,l,p,q) \in D_{4,N}} \frac{y_ky_ly_py_q}{ \pi_k \pi_l \pi_p \pi_q}|\mathbb{E}[(I_{k}-\pi_{k})(I_{l}-\pi_{l})(I_{p}-\pi_{p})(I_{q}-\pi_{q}) ]| \\
		&\leq \frac{1}{\lambda^4}\frac{\sum_{k \in U} y_k^4 }{N} \left(\frac{2n_N}{N^2}+\frac{2n_N}{N} \underset{(k,p,q) \in D_{3,N}}{\max}|\mathbb{E}[(I_{k}-\pi_{k})^2(I_{p}-\pi_{p})(I_{q}-\pi_{q}) ]| \right. \\ &+ \left. n_N \underset{(k,l,p,q) \in D_{4,N}}{\max}|\mathbb{E}[(I_{k}-\pi_{k})(I_{l}-\pi_{l})(I_{p}-\pi_{p})(I_{q}-\pi_{q}) ]| \right), 
		\end{align*}
		where the last term diminishes as $N \rightarrow \infty$ by Assumptions (A7)-(A8). This concludes the proof.
	\end{proof}
	
	\textbf{Lemma 2.} 
	Let $m\in \mathbb{N}$. Assume that $X_i-Y_i=O_p\left(a_n\right)$ for $i=1,2, \dots, m$. Then,
	\begin{align*}
	f\left(X_1,X_2,\dots,X_m \right)-f\left(Y_1,Y_2,\dots,Y_m \right)&=O_p\left(a_n\right)
	\end{align*}
	where $f(\cdot)$ could be either $\min(\cdot)$ or $\max(\cdot)$ coordinate-wise function. 
	
	\begin{proof}[Proof of Lemma 2]
		We are going to prove this proposition by induction in $m$. The case $m=1$ is clear since $f(X_1)-f(Y_1)=X_1-Y_1=O_p(a_n)$. Assume the result is true for $m=k$. That is,
		\begin{align*}
		f\left(X_1,X_2,\dots,X_k \right)-f\left(Y_1,Y_2,\dots,Y_k \right)&=O_p\left(a_n\right).
		\end{align*}
		
		We need to prove that the result is true for $m=k+1$. Note that 
		\begin{align*}
		f(X_1,\dots, X_k,X_{k+1})&=f(f(X_1,\dots, X_k),X_{k+1}),
		\end{align*}
		which is also true for the sequence of $Y$'s. 
		
		Denote $u_k=f(X_1,\dots, X_k)$ and $v_k=f(Y_1,\dots, Y_k)$. By the induction assumption, $u_k-v_k=O_p(a_n).$ For the rest of the proof we are going to consider only the case when $f(\cdot)=\min(\cdot)$. Later we will note that the proof for $f(\cdot)=\max(\cdot)$ is analogous to what follows.
		
		Note that we can write $\min(u_k,X_{k+1})=\frac{1}{2} \left(u_k+X_{k+1}-|u_k-X_{k+1}| \right)$ and $\min(v_k,Y_{k+1})=\frac{1}{2} \left(v_k+Y_{k+1}-|v_k-Y_{k+1}| \right)$. Hence,
		\begin{align*}
		&\left|\min(u_k,X_{k+1})-\min(v_k,Y_{k+1})\right|\\
		&=\frac{1}{2}\left|(u_k-v_k)+(X_{k+1}-Y_{k+1})+\left(|v_k-Y_{k+1}|-|u_k-X_{k+1}|\right)\right|\\
		&\leq\frac{1}{2}\left\{\left|u_k-v_k\right|+\left|X_{k+1}-Y_{k+1}\right|+\left||v_k-Y_{k+1}|-|u_k-X_{k+1}|\right| \right\}\\
		&\leq \frac{1}{2}\left\{\left|u_k-v_k\right|+\left|X_{k+1}-Y_{k+1}\right|+\left|(v_k-u_k)-(X_{k+1}-Y_{k+1}\right| \right\}\\
		&\leq \frac{1}{2}\left\{\left|u_k-v_k\right|+\left|X_{k+1}-Y_{k+1}\right|+|v_k-u_k|+|X_{k+1}-Y_{k+1}| \right\}\\
		&=\left|u_k-v_k\right|+\left|X_{k+1}-Y_{k+1}\right|.
		\end{align*}
		Since both $u_k-v_k=O_p(a_n)$ and $X_{k+1}-Y_{k+1}=O_p(a_n)$, then for any $\epsilon>0$ there exist $\delta_1>0$ and $\delta_2>0$ such that
		\begin{align*}
		P(a_n^{-1}|u_k-v_k|>\delta_1)<\frac{\epsilon}{2} \; \; &\text{ and } \; \;  P(a_n^{-1}|X_{k+1}-Y_{k+1}|>\delta_2)<\frac{\epsilon}{2}.
		\end{align*}
		Therefore,
		\begin{align*}
		\epsilon=\frac{\epsilon}{2}+\frac{\epsilon}{2}&>P(a_n^{-1}|u_k-v_k|>\delta_1)+P(a_n^{-1}|X_{k+1}-Y_{k+1}|>\delta_2)\\
		&\geq P\left(a_n^{-1}\left|u_k-v_k\right|+a_n^{-1}\left|X_{k+1}-Y_{k+1}\right|>\delta_1+\delta_2\right) \\
		& \geq P\left( a_n^{-1}\left|\min(u_k,X_{k+1})-\min(v_k,Y_{k+1})\right| >\delta_1+\delta_2\right).
		\end{align*}
		Setting $\delta^{*}=\delta_1+\delta_2$, then we can conclude that $\min(u_k,X_{k+1})-\min(v_k,Y_{k+1})=O_p(a_n)$. Thus, the result is true for $m=k+1$. For the case when $f(\cdot)=\max(\cdot)$, we just need to use the fact that $\max(u_k,X_{k+1})=\frac{1}{2} \left(u_k+X_{k+1}+|u_k-X_{k+1}| \right)$ and then follow an analogous proof as above.
	\end{proof}	
	
	\textbf{Lemma 3.} Let $\boldsymbol{\theta}_{\mu}=\left( \theta_{\mu_1}, \theta_{\mu_2}, \dots, \theta_{\mu_D}\right)^{\top}$ be the weighted isotonic vector of the limiting domain means $\boldsymbol{\mu}$ with weights $\gamma_1,\gamma_2,\dots, \gamma_D$. Then, 
	$$ \tilde{\theta}_{s_d}-\theta_{\mu_d}=O_p(n_N^{-1/2} ), \; \; \; \textsl{ for } d=1,2, \dots, D. $$ 
	
	\begin{proof}[Proof of Lemma 3]
		Fix $d$. Following the proof of Lemma 2, it can be proved that $\theta_{U_d}-\theta_{\mu_d}=O\left(N^{-1/2}\right)$ from Assumption (A4). By Theorem 4, $\tilde{\theta}_{s_d}-\theta_{U_d}=O_p( n_N^{-1/2})$. Therefore, we can conclude that $\tilde{\theta}_{s_d}-\theta_{\mu_d}=O_p(n_N^{-1/2} )$.
	\end{proof}
	
	\textbf{Lemma 4.} $(\boldsymbol{\tilde{y}}_{s}-\boldsymbol{\tilde{\theta}}_s)^{\top} \boldsymbol{W}_s (\boldsymbol{\tilde{y}}_{s}-\boldsymbol{\tilde{\theta}}_s)=\left(\boldsymbol{\mu}-\boldsymbol{\theta_{\mu}}\right)^{\top}\boldsymbol{\Gamma}\left(\boldsymbol{\mu}-\boldsymbol{\theta_{\mu}}\right) + O_p(n_N^{1/2})$, where $\boldsymbol{\Gamma}=\text{diag}(\gamma_1,\gamma_2, \dots, \gamma_D)$. 
	
	\begin{proof}[Proof of Lemma 4]
		From $\boldsymbol{\tilde{y}}_s-\overline{\boldsymbol{y}}_U=\boldsymbol{1} O_p\left(N^{-1/2}\right)$ and $\overline{\boldsymbol{y}}_U-\boldsymbol{\mu}=\boldsymbol{1} O\left(N^{-1/2}\right)$, we get that $ \boldsymbol{\tilde{y}}_s-\boldsymbol{\mu}=\boldsymbol{1} O_p(n_N^{-1/2})$. Further,  $\boldsymbol{\tilde{\theta}}_s-\boldsymbol{\theta_{\mu}}=\boldsymbol{1} O_p(n_N^{-1/2})$ by Lemma 3. Therefore, $\boldsymbol{\tilde{y}}_{s}-\boldsymbol{\tilde{\theta}}_s=\boldsymbol{\mu}-\boldsymbol{\theta}_{\boldsymbol{\mu}}+\boldsymbol{1}O_p(n_N^{-1/2})$. In addition, $\widehat{N}_d/\widehat{N}=\gamma_d+O_p(n_N^{-1/2})$ for $d=1,\dots, D$. Thus, $(\boldsymbol{\tilde{y}}_{s}-\boldsymbol{\tilde{\theta}}_s)^{\top} \boldsymbol{W}_s (\boldsymbol{\tilde{y}}_{s}-\boldsymbol{\tilde{\theta}}_s)=\left(\boldsymbol{\mu}-\boldsymbol{\theta_{\mu}}\right)^{\top}\boldsymbol{\Gamma}\left(\boldsymbol{\mu}-\boldsymbol{\theta_{\mu}}\right)+O_p(n^{-1/2})$.	
	\end{proof}
	
	\textbf{Lemma 5.}
	$\text{cov}(\widehat{\theta}_{s_i},\widehat{\theta}_{s_j})=O(n_N^{-1})$, for any $i,j=1,2,\dots, D$.
	
	\begin{proof}[Proof of Lemma 5]
		Define $\mathbb{F}$ to the set of representative elements $F_i$, and $\boldsymbol{P}_{F_i}$ as it was done in the proof of Theorem 2. In addition, let $\mathbb{F}_1$ be the set of representative elements $F_i$ of those poolings that correspond to the disjoint sets $J^0$ and $J^1$ such that $J^{0}_{\mu} \subseteq J^0$ and $J^{1}_{\mu} \subseteq J^1$. That is, $F_i \in \mathbb{F}_1$ if and only if the pooling represented by $F_i$ is allowed by $\boldsymbol{\mu}$ to produce $\boldsymbol{\theta}_{\mu}$. Further, let $\mathbb{F}_2=\mathbb{F} \setminus \mathbb{F}_1$. 
		
		Suppose that there exist indexes $i \neq j$ such that $F_i, F_j \in \mathbb{F}_1$. First, note that both $\boldsymbol{P}_{F_i}\boldsymbol{\overline{y}}_U$ and $\boldsymbol{P}_{F_j}\boldsymbol{\overline{y}}_U$ converge to the vector $\boldsymbol{\theta}_{\mu}$. From Assumption (A4), $\boldsymbol{P}_{F_i}\boldsymbol{\overline{y}}_U-\boldsymbol{\theta}_{\mu}=\boldsymbol{1}O(N^{-1/2})$ and $\boldsymbol{P}_{F_j}\boldsymbol{\overline{y}}_U-\boldsymbol{\theta}_{\mu}=\boldsymbol{1}O(N^{-1/2})$, which implies that $\boldsymbol{P}_{F_i}\boldsymbol{\overline{y}}_U-\boldsymbol{P}_{F_j}\boldsymbol{\overline{y}}_U=\boldsymbol{1}O(N^{-1/2})$.
		
		Consider any index $i$ such that $F_i \in \mathbb{F}_1$. Denote $p_{i,kl}$ to the $(k,l)$-element of $\boldsymbol{P}_{F_i}$. Fix $d$. From the fact that the function $\mathbb{E}[(\widehat{\theta}_{s_d}-x)^2]$ is minimized by the constant $x=\mathbb{E}(\widehat{\theta}_{s_d})$, then we have
		\begin{align*}
		\text{var}(\widehat{\theta}_{s_d}) &= \mathbb{E}\left\{\left[\widehat{\theta}_{s_d}-\mathbb{E}(\widehat{\theta}_{s_d})\right]^2\right\} \\
		& \leq \mathbb{E}\left\{\left[\widehat{\theta}_{s_d}-\left(\sum_{j=1}^{D} p_{i,dj}\overline{y}_{U_j} \right) \right]^2\right\}\\
		&= \mathbb{E}\left\{\left[ \left( \sum_{k=1}^{|\mathbb{F}|} \widehat{\theta}_{s_d}I\{\boldsymbol{\widehat{y}}_{s} \in F_k \} \right) - \sum_{k=1}^{|\mathbb{F}|}\left(\sum_{j=1}^{D} p_{i,dj}\overline{y}_{U_j} \right)I\{\boldsymbol{\widehat{y}}_{s} \in F_k \} \right]^2\right\} \\
		&= \mathbb{E}\left\{\left[ \sum_{k=1}^{|\mathbb{F}|} \left(\sum_{j=1}^{D} p_{k,dj}\widehat{y}_{s_j}-\sum_{j=1}^{D} p_{i,dj}\overline{y}_{U_j} \right)I\{\boldsymbol{\widehat{y}}_{s} \in F_k \} \right]^2\right\} \\
		& \leq |\mathbb{F}| \sum_{k=1}^{|\mathbb{F}|} \mathbb{E} \left[  \left(\sum_{j=1}^{D} p_{k,dj}\widehat{y}_{s_j} -\sum_{j=1}^{D} p_{i,dj}\overline{y}_{U_j} \right)^2 I\{\boldsymbol{\widehat{y}}_{s} \in F_k \} \right] \\
		& = |\mathbb{F}| \left\{ \sum_{F_k \in |\mathbb{F}_1|} \mathbb{E} \left[  \left(\sum_{j=1}^{D} p_{k,dj}\widehat{y}_{s_j} -\sum_{j=1}^{D} p_{i,dj}\overline{y}_{U_j} \right)^2 I\{\boldsymbol{\widehat{y}}_{s} \in F_k \} \right] \right. \\
		& \left. + \sum_{F_k \in |\mathbb{F}_2|} \mathbb{E} \left[  \left(\sum_{j=1}^{D} p_{k,dj}\widehat{y}_{s_j} -\sum_{j=1}^{D} p_{i,dj}\overline{y}_{U_j} \right)^2 I\{\boldsymbol{\widehat{y}}_{s} \in F_k \} \right]\right\} \\
		& \leq |\mathbb{F}| \sum_{F_k \in |\mathbb{F}_1|} \mathbb{E} \left[  \left(\sum_{j=1}^{D} p_{k,dj}\widehat{y}_{s_j} -\sum_{j=1}^{D} p_{i,dj}\overline{y}_{U_j} \right) ^2 \right] + o(n_N^{-1}) \\
		&= |\mathbb{F}| \sum_{F_k \in |\mathbb{F}_1|} \left[ \text{var} \left( \sum_{j=1}^{D} p_{k,dj}\widehat{y}_{s_j} \right) +    \left(\sum_{j=1}^{D} p_{k,dj}\overline{y}_{U_j} -\sum_{j=1}^{D} p_{i,dj}\overline{y}_{U_j} \right) ^2 \right] + o(n_N^{-1}) \\
		&= |\mathbb{F}| \left[ \sum_{F_k \in |\mathbb{F}_1|} \text{var} \left( \sum_{j=1}^{D} p_{k,dj}\widehat{y}_{s_j} \right) + \sum_{F_k \in |\mathbb{F}_1|} \left(\sum_{j=1}^{D} p_{k,dj}\overline{y}_{U_j} -\sum_{j=1}^{D} p_{i,dj}\overline{y}_{U_j} \right) ^2 \right] + o(n_N^{-1}) \\
		&= O(n_N^{-1}) + O(N^{-1}) + o(n_N^{-1}),
		\end{align*}
		which implies that $\text{var}(\widehat{\theta}_{s_d})=O(n_N^{-1})$. Thus, by the Cauchy-Schwartz inequality, we conclude that $\text{cov}(\widehat{\theta}_{s_i},\widehat{\theta}_{s_j})=O(n_N^{-1})$ for $i,j=1,2,\dots, D$. 
	\end{proof}

	\begin{proof}[Proof of Proposition 1]
		Note that
		\begin{align*}
		\text{PSE}(\boldsymbol{\widehat{\theta}}_s)&=\mathbb{E}\left[(\boldsymbol{\widehat{y}}_{s^*}-\boldsymbol{\widehat{\theta}}_{s})^{\top} \boldsymbol{W}_U (\boldsymbol{\widehat{y}}_{s^*}-\boldsymbol{\widehat{\theta}}_{s})\right]\\
		&=\mathbb{E} \left[(\boldsymbol{\widehat{y}}_{s^*}-\boldsymbol{\overline{y}}_U)^{\top} \boldsymbol{W}_U (\boldsymbol{\widehat{y}}_{s^*}-\boldsymbol{\overline{y}}_U) \right]+2 \mathbb{E} \left[(\boldsymbol{\overline{y}}_{s^*}-\boldsymbol{\overline{y}}_U)^{\top} \boldsymbol{W}_U (\boldsymbol{\overline{y}}_{U}-\boldsymbol{\widehat{\theta}}_{s}) \right] \\
		&+\mathbb{E} \left[(\boldsymbol{\overline{y}}_{U}-\boldsymbol{\widehat{\theta}}_{s})^{\top} \boldsymbol{W}_U (\boldsymbol{\overline{y}}_{U}-\boldsymbol{\widehat{\theta}}_{s}) \right]\\
		&= \Tr \left[\boldsymbol{W}_U \text{cov} ( \boldsymbol{\widehat{y}}_s,\boldsymbol{\widehat{y}}_s) \right]+\mathbb{E} \left[(\boldsymbol{\overline{y}}_{U}-\boldsymbol{\widehat{\theta}}_{s})^{\top} \boldsymbol{W}_U (\boldsymbol{\overline{y}}_{U}-\boldsymbol{\widehat{\theta}}_{s} ) \right].
		\end{align*}
		By adding and subtracting $\boldsymbol{\widehat{y}}_s$ in the expectation term of the above equality, we have that
		\begin{align*}
		\mathbb{E} \left[(\boldsymbol{\overline{y}}_{U}-\boldsymbol{\widehat{\theta}}_{s})^{\top} \boldsymbol{W}_U (\boldsymbol{\overline{y}}_{U}-\boldsymbol{\widehat{\theta}}_{s}) \right]
		&=\Tr \left[\boldsymbol{W}_U \text{cov} ( \boldsymbol{\widehat{y}}_s,\boldsymbol{\widehat{y}}_s) \right]\\
		&+2\mathbb{E} \left[(\boldsymbol{\overline{y}}_{U}-\boldsymbol{\widehat{y}}_s)^{\top} \boldsymbol{W}_U (\boldsymbol{\widehat{y}}_{s}-\boldsymbol{\widehat{\theta}}_{s}) \right]+\mathbb{E}\left[\text{SSE}(\boldsymbol{\widehat{\theta}}_s)\right].
		\end{align*}
		Further,
		\begin{align*}
		\mathbb{E} \left[(\boldsymbol{\overline{y}}_{U}-\boldsymbol{\widehat{y}}_s)^{\top} \boldsymbol{W}_U (\boldsymbol{\widehat{y}}_{s}-\boldsymbol{\widehat{\theta}}_{s}) \right]&=\mathbb{E} \left[(\boldsymbol{\overline{y}}_{U}-\boldsymbol{\widehat{y}}_s)^{\top} \boldsymbol{W}_U \boldsymbol{\widehat{y}}_{s} \right]+\mathbb{E}\left[(\boldsymbol{\widehat{y}}_{s}-\boldsymbol{\overline{y}}_U)^{\top} \boldsymbol{W}_U\boldsymbol{\widehat{\theta}}_{s} \right]\\
		&=-\Tr \left[\boldsymbol{W}_U \text{cov} ( \boldsymbol{\widehat{y}}_s,\boldsymbol{\widehat{y}}_s) \right] + \Tr \left[ \boldsymbol{W}_U \text{cov} ( \boldsymbol{\widehat{\theta}}_{s},\boldsymbol{\widehat{y}}_s)  \right].
		\end{align*}
		Hence, $\text{PSE}(\boldsymbol{\widehat{\theta}}_s)=\mathbb{E}\left[\text{SSE}(\boldsymbol{\widehat{\theta}}_s)\right]+ 2 \Tr \left[ \boldsymbol{W}_U \text{cov} ( \boldsymbol{\widehat{\theta}}_{s},\boldsymbol{\widehat{y}}_s) \right]$. 
	\end{proof}

	\begin{proof}[Proof of Theorem 1]
		First, consider an index $i$ such that $i \in J_{\mu}^0$ and assume that $i \notin J_s^0$. Define $L_{\mu}=J_{\mu}^1 \cup \{0,D\}$. Consider the largest index $l\in L_{\mu}$ that is less than $i$, and the smallest index $u \in L_{\mu}$ that is greater than $i$. Then, the slope from point $G_{\mu}(l)$ to $G_{\mu}(i)$ is greater than the slope from point $G_{\mu}(i)$ to $G_{\mu}(u)$. That is, $\mu_{l+1:i} > \mu_{i+1:u}$. Now, since $i \notin J_s^0$, then the slope from point $G_s(l)$ to $G_s(i)$ is at most equal to the slope from point $G_s(i)$ to $G_s(u)$. That implies $\widehat{y}_{s_{l+1:i}} \leq \widehat{y}_{s_{i+1:u}}$. Therefore, we have
		\begin{align*}
		P(i \notin J_s^0)&=P\left( \widehat{y}_{s_{i+1:u}} \geq \widehat{y}_{s_{l+1:i}} \right )\\
		&=P\left( (\widehat{y}_{s_{i+1:u}}-\mu_{i+1:u}) - (\widehat{y}_{s_{l+1:i}} -\mu_{l+1:i}) \geq \mu_{l+1:i}-\mu_{i+1:u} \right) \\
		&\leq \frac{ \mathbb{E}\left\{[(\widehat{y}_{s_{i+1:u}}-\mu_{i+1:u}) - (\widehat{y}_{s_{l+1:i}} -\mu_{l+1:i})]^4\right\} }{ (\mu_{l+1:i}-\mu_{i+1:u})^4 } = o(n_N^{-1}),
		\end{align*}  
		where the last equality comes from Lemma 1 and Assumption (A4). Thus, $P(A_0^c)=o(n^{-1})$.
		
		Now, consider an index $i$ such that $i \in J_{\mu}^1$ but $i \notin J_s^1$. Let $L_s=J_s^1\cup \{0,D\}$. Let $l,u \in L_{s}$ be the largest index less than $i$ and the smallest index greater than $i$, respectively. Since $i$ is not a corner point of $G_s$, then $G_s(i)$ is either on or above it, i.e. $\widehat{y}_{s_{l+1:i}} \geq \widehat{y}_{s_{i+1:u}}$. Moreover, $\mu_{l+1:i} < \mu_{i+1:u} $ because $i$ is a corner point of $G_{\mu}$. Hence,
		\begin{align*}
		P(i \notin J_s^1)&=P\left( \widehat{y}_{s_{l+1:i}} \geq \widehat{y}_{s_{i+1:u}} \right )\\
		&=P\left( (\widehat{y}_{s_{l+1:i}}-\mu_{l+1:i}) - (\widehat{y}_{s_{i+1:u}} -\mu_{i+1:u}) \geq \mu_{i+1:u}-\mu_{l+1:i} \right) \\
		&\leq \frac{ \mathbb{E}\left\{[(\widehat{y}_{s_{l+1:i}}-\mu_{l+1:i}) - (\widehat{y}_{s_{i+1:u}} -\mu_{i+1:u})]^4\right\} }{ (\mu_{i+1:u}-\mu_{l+1:i})^4 } = o(n_N^{-1}),
		\end{align*} 
		which leads to the conclusion that $P(A_1^c)=o(n_N^{-1})$. 	
	\end{proof}
	
	
	\begin{proof}[Proof of Theorem 2]
		Let $F_1, F_2, \dots, F_{2^{D-1}}$ be representative elements for each of the possible poolings (groupings) for a vector of length $D$. Also, define $\mathbb{F}$ to the set of all of these representative elements. Since $J_\mu^0 \cup J_\mu^1 = \{1, 2, \dots, D-1\}$ and without loss of generality, let $F_1$ be the representative element of the unique pooling allowed by  $\boldsymbol{\mu}$. Denote $\boldsymbol{P}_{F_i}$ to be the weighted projection matrix that corresponds to the pooling represented by $F_i$ with weights $N_1, N_2, \dots, N_D$. Also, define $P(\boldsymbol{\widehat{y}}_s \in F_i)$ to the probability that the pooling represented by $F_i$ is allowed by $\boldsymbol{\widehat{y}}_s$ to obtain $\boldsymbol{\widehat{\theta}}_s$. By Theorem 1,
		\begin{center}
			P$(\boldsymbol{\widehat{y}}_s \in F_i) = \left\{
			\begin{array}{ll}
			1+o(n_N^{-1}), & \text{ if }  i=1; \\
			o(n_N^{-1}), & \text{ if } i \neq 1.
			\end{array} \right. $
		\end{center}
		Also, since $|\widehat{y}_{s_d}| \leq \lambda^{-1} N_d^{-1} \sum_{k \in U_d}|y_k|$ for $d=1,\dots, D$, then for $i \neq 1$,
		\begin{align*}
		\left|\mathbb{E}(\widehat{y}_{s_d} I\{\boldsymbol{\widehat{y}}_s \in F_i\})\right| &\leq \mathbb{E}(|\widehat{y}_{s_d}| I\{\boldsymbol{\widehat{y}}_s \in F_i\}) \\
		&\leq \left(  \frac{1}{\lambda N_d} \sum_{k \in U_d}|y_k| \right) P(\boldsymbol{\widehat{y}}_s \in F_i)\\
		&\leq \lambda^{-1} \left(  \frac{1}{N_d} \sum_{k \in U_d}y_k^4 \right)^{1/4} P(\boldsymbol{\widehat{y}}_s \in F_i)=o(n_N^{-1}),
		\end{align*}
		which implies that $\mathbb{E}(\boldsymbol{\widehat{y}}_s I\{\boldsymbol{\widehat{y}}_s \in F_i\} )=\boldsymbol{1}o(n_N^{-1})$. Hence,
		\begin{align*}
		\mathbb{E}(\boldsymbol{\widehat{y}}_s)=\sum_{i=1}^{|\mathbb{F}|}\mathbb{E}(\boldsymbol{\widehat{y}}_s I\{ \boldsymbol{\widehat{y}}_s \in F_i\})=\mathbb{E}(\boldsymbol{\widehat{y}}_s I\{ \boldsymbol{\widehat{y}}_s \in F_1\})+\boldsymbol{1}o({n_N^{-1}}).
		\end{align*}
		Then, we obtain that
		\begin{align*}
		\mathbb{E}(\boldsymbol{\widehat{\theta}}_s)&=\sum_{i=1}^{|\mathbb{F}|}\mathbb{E}(\boldsymbol{\widehat{\theta}}_s I\{ \boldsymbol{\widehat{y}}_s \in F_i\})=\sum_{i=1}^{|\mathbb{F}|}\mathbb{E}(\boldsymbol{P}_{F_i}\boldsymbol{\widehat{y}}_s I\{ \boldsymbol{\widehat{y}}_s \in F_i\})\\
		&=\boldsymbol{P}_{F_1}\mathbb{E}(\boldsymbol{\widehat{y}}_s I\{ \boldsymbol{\widehat{y}}_s \in F_1\})+\boldsymbol{1}o({n_N^{-1}})=\boldsymbol{P}_{F_1}\mathbb{E}(\boldsymbol{\widehat{y}}_s)+\boldsymbol{1}o({n_N^{-1}}).
		\end{align*}
		Analogously, $\mathbb{E}(\boldsymbol{\widehat{\theta}}_s\boldsymbol{\widehat{y}}_s^{\top})=\boldsymbol{P}_{F_1}\mathbb{E}(\boldsymbol{\widehat{y}}_s\boldsymbol{\widehat{y}}_s^{\top})+\boldsymbol{J}o(n_N^{-1})$, where $\boldsymbol{J}$ is the $D \times D$ matrix of ones. Therefore, we can conclude that	
		\begin{align*}
		\text{cov}(\boldsymbol{\widehat{\theta}}_s, \boldsymbol{\widehat{y}}_s)&=\mathbb{E}(\boldsymbol{\widehat{\theta}}_s\boldsymbol{\widehat{y}}_s^{\top})-\mathbb{E}(\boldsymbol{\widehat{\theta}}_s)\mathbb{E}(\boldsymbol{\widehat{y}}_s)^{\top}\\
		&=\boldsymbol{P}_{F_1}\mathbb{E}(\boldsymbol{\widehat{y}}_s\boldsymbol{\widehat{y}}_s^{\top})-\boldsymbol{P}_{F_1}\mathbb{E}(\boldsymbol{\widehat{y}}_s)\mathbb{E}(\boldsymbol{\widehat{y}}_s)^{\top} + \boldsymbol{J}o(n_N^{-1})\\ 
		&=\boldsymbol{P}_{F_1}[\mathbb{E}(\boldsymbol{\widehat{y}}_s\boldsymbol{\widehat{y}}_s^{\top})-\mathbb{E}(\boldsymbol{\widehat{y}}_s)\mathbb{E}(\boldsymbol{\widehat{y}}_s)^{\top}]+ \boldsymbol{J}o(n_N^{-1})\\
		&=\boldsymbol{P}_{F_1}\text{var}(\boldsymbol{\widehat{y}}_s) + \boldsymbol{J}o(n_N^{-1}).
		\end{align*}
		
		Now, note that
		\begin{align*}
		\left|\widehat{\Sigma}_{dd} \right| \leq \frac{1}{\lambda^2}\frac{\sum_{k \in U_d} y_k^2}{N_d} \left( \frac{1}{N_d} + 1  \right) \leq  \frac{1}{\lambda^2} \left( \frac{\sum_{k \in U_d} y_k^4}{N_d} \right)^{1/2} \left( \frac{1}{N_d} + 1  \right), 
		\end{align*}
		which implies that $\mathbb{E}( \boldsymbol{\widehat{\Sigma}} I\{ \boldsymbol{\widehat{y}}_s \in F_i \})=\boldsymbol{J}o(n_N^{-1})$ for $i \neq 1$. Moreover,
		\begin{align*}
		\mathbb{E}(\boldsymbol{\widehat{\Sigma}})=\sum_{i=1}^{|\mathbb{F}|}\mathbb{E}(\boldsymbol{\widehat{\Sigma}} I\{ \boldsymbol{\widehat{y}}_s \in F_i\})=\mathbb{E}(\boldsymbol{\widehat{\Sigma}} I\{ \boldsymbol{\widehat{y}}_s \in F_1\})+\boldsymbol{J}o({n_N^{-1}}).
		\end{align*}
		Then,
		\begin{align*}
		\mathbb{E}(\boldsymbol{\widehat{P}}_s \boldsymbol{\widehat{\Sigma}}) &= \sum_{i=1}^{|\mathbb{F}|} \mathbb{E}(\boldsymbol{P}_{F_i} \boldsymbol{\widehat{\Sigma}} I\{\boldsymbol{\widehat{y}}_s \in F_i\}) = \sum_{i=1}^{|\mathbb{F}|} \boldsymbol{P}_{F_i}\mathbb{E}(\boldsymbol{\widehat{\Sigma}} I\{\boldsymbol{\widehat{y}}_s \in F_i\})\\
		&= \boldsymbol{P}_{F_1}\mathbb{E}(\boldsymbol{\widehat{\Sigma}} I\{\boldsymbol{\widehat{y}}_s \in F_1\})+ \boldsymbol{J}o(n_N^{-1}) = \boldsymbol{P}_{F_1}\mathbb{E}(\boldsymbol{\widehat{\Sigma}})+\boldsymbol{J}o(n_N^{-1})\\
		&=\boldsymbol{P}_{F_1}\text{var}(\boldsymbol{\widehat{y}}_{s})+\boldsymbol{J}o(n_N^{-1}).
		\end{align*}
		Thus, from Proposition 1,
		\begin{align*}
		\mathbb{E}[\text{CIC}_s(\boldsymbol{\widehat{\theta}}_s)]-\text{PSE}(\boldsymbol{\widehat{\theta}}_s)=2\Tr\{\boldsymbol{W}_U[\mathbb{E}(\boldsymbol{\widehat{P}}_s\boldsymbol{\widehat{\Sigma}} ) - \text{cov}(\boldsymbol{\widehat{\theta}}_s, \boldsymbol{\widehat{y}}_s)]\}=o(n_N^{-1}).
		\end{align*}
	\end{proof}

	\begin{proof}[Proof of Theorem 3]
		The $AC(\tilde{y}_{s_{i_1:j_1}},\tilde{y}_{s_{i_2:j_2}})$ term can be broken into two sums: one with the common and one with the uncommon elements of $U_{i_1:j_1}$ and $U_{i_2:j_2}$. By doing that, we get
		\begin{align*}
		n_N \left| AC(\tilde{y}_{s_{i_1:j_1}},\tilde{y}_{s_{i_2:j_2}})\right| &= \frac{n_N}{N_{i_1:j_1}N_{i_2:j_2}}\left|\underset{k \in U_{i_1:j_1}}{\sum} \underset{l \in U_{i_2:j_2}}{\sum} \Delta_{kl} \left(\frac{y_k-\overline{y}_{U_{i_1:j_1}}}{\pi_k}\right)\left(\frac{y_l-\overline{y}_{U_{i_2:j_2}}}{\pi_l}\right) \right|\\
		&\leq \frac{n_N}{N_{i_1:j_1}N_{i_2:j_2}}\left|\underset{k \in U_{i_1:j_1} \cap U_{i_2:j_2} }{\sum} \frac{1-\pi_k}{\pi_k} \left(y_k-\overline{y}_{U_{i_1:j_1}}\right) \left(y_k-\overline{y}_{U_{i_2:j_2}}\right) \right| \\
		&+ \frac{n_N}{N_{i_1:j_1}N_{i_2:j_2}}\left|\underset{k\neq l}{\underset{k \in U_{i_1:j_1}}{\sum} \underset{l \in U_{i_2:j_2}}{\sum} } \Delta_{kl} \left(\frac{y_k-\overline{y}_{U_{i_1:j_1}}}{\pi_k}\right)\left(\frac{y_l-\overline{y}_{U_{i_2:j_2}}}{\pi_l}\right) \right| \\ 
		&\leq \frac{n_N}
		{N\lambda}\frac{N^2}{N_{i_1:j_1}N_{i_2:j_2}} \left( \frac{\underset{k \in U_{i_1:j_1} \cap U_{i_2:j_2}}{\sum}  \left(y_k-\overline{y}_{U_{i_1:j_1}}\right)^2}{N} \right. \\
		&+ \left. \frac{\underset{k \in U_{i_1:j_1} \cap U_{i_2:j_2}}{\sum} \left(y_k-\overline{y}_{U_{i_2:j_2}}\right)^2}{N} \right) \\
		&+ \frac{n_N \underset{k,l \in U_N: \; k\neq l}{\max}|\Delta_{kl}|}{\lambda^2} \left( \frac{\underset{k \in U_{i_1:j_1} }{\sum} \left(y_k-\overline{y}_{U_{i_1:j_1}}\right)^2}{N_{i_1:j_1}} + \frac{\underset{l \in U_{i_2:j_2} }{\sum} \left(y_l-\overline{y}_{U_{i_2:j_2}}\right)^2}{N_{i_2:j_2}} \right),
		\end{align*}
		where the last inequality is obtained from Assumption (A5). Given that each of the terms in the above upper bound is asymptotically bounded by Assumptions (A2)-(A5), then the first result is true. 
		
		To show the second result, note that
		\begin{align*}
		& n_N\left|\frac{\widehat{N}_{i_1:j_1}\widehat{N}_{i_2:j_2}}{N_{i_1:j_1}N_{i_2:j_2}}\widehat{AC}(\tilde{y}_{s_{i_1:j_1}},\tilde{y}_{s_{i_2:j_2}}) - AC(\tilde{y}_{s_{i_1:j_1}},\tilde{y}_{s_{i_2:j_2}})\right| \\
		&=\frac{n_N}{N_{i_1:j_1}N_{i_2:j_2}} \left| \underset{k \in U_{i_1:j_1}}{\sum} \underset{l \in U_{i_2:j_2}}{\sum} \Delta_{kl} \left(\frac{y_k-\tilde{y}_{s_{i_1:j_1}}}{\pi_k}\right)\left(\frac{y_l-\tilde{y}_{s_{i_2:j_2}}}{\pi_l}\right)\frac{I_kI_l}{\pi_{kl}} \right. \\
		&- \left. \underset{k \in U_{i_1:j_1}}{\sum} \underset{l \in U_{i_2:j_2}}{\sum} \Delta_{kl} \left(\frac{y_k-\overline{y}_{U_{i_1:j_1}}}{\pi_k}\right)\left(\frac{y_l-\overline{y}_{U_{i_2:j_2}}}{\pi_l}\right) \right| \\
		& \leq \frac{n_N}{N_{i_1:j_1}N_{i_2:j_2}} \left| \underset{k \in U_{i_1:j_1}}{\sum} \underset{l \in U_{i_2:j_2}}{\sum} \Delta_{kl} \left(\frac{y_k-\overline{y}_{U_{i_1:j_1}}}{\pi_k}\right)\left(\frac{y_l-\overline{y}_{U_{i_2:j_2}}}{\pi_l}\right) \left( \frac{I_kI_l- \pi_{kl}}{\pi_{kl}} \right)\right|\\
		&+ \frac{n_N}{N_{i_1:j_1}N_{i_2:j_2}} \left| \underset{k \in U_{i_1:j_1}}{\sum} \underset{l \in U_{i_2:j_2}}{\sum} \Delta_{kl} \left(\frac{y_k-\overline{y}_{U_{i_1:j_1}}}{\pi_k}\right)\left(\frac{\overline{y}_{U_{i_2:j_2}}-\tilde{y}_{s_{i_2:j_2}}}{\pi_l}\right)  \frac{I_kI_l}{\pi_{kl}} \right|\\
		&+ \frac{n_N}{N_{i_1:j_1}N_{i_2:j_2}} \left| \underset{k \in U_{i_1:j_1}}{\sum} \underset{l \in U_{i_2:j_2}}{\sum} \Delta_{kl} \left(\frac{\overline{y}_{U_{i_1:j_1}}-\tilde{y}_{s_{i_1:j_1}}}{\pi_k}\right)\left(\frac{y_l-\overline{y}_{U_{i_2:j_2}}}{\pi_l}\right)  \frac{I_kI_l}{\pi_{kl}} \right|\\
		&+ \frac{n_N}{N_{i_1:j_1}N_{i_2:j_2}} \left| \underset{k \in U_{i_1:j_1}}{\sum} \underset{l \in U_{i_2:j_2}}{\sum} \Delta_{kl} \left(\frac{\overline{y}_{U_{i_1:j_1}}-\tilde{y}_{s_{i_1:j_1}}}{\pi_k}\right)\left(\frac{\overline{y}_{U_{i_2:j_2}}-\tilde{y}_{s_{i_2:j_2}}}{\pi_l}\right)  \frac{I_kI_l}{\pi_{kl}} \right|\\
		&=a_{1N}+a_{2N}+a_{3N}+a_{4N},
		\end{align*}
		where we used the identities $y_k-\tilde{y}_{s_{i_1:j_1}}=\left(y_k-\overline{y}_{U_{i_1:j_1}} \right)+ \left( \overline{y}_{U_{i_1:j_1}}-\tilde{y}_{s_{i_1:j_1}}\right)$, and $y_l-\tilde{y}_{s_{i_2:j_2}}=\left(y_l-\overline{y}_{U_{i_2:j_2}} \right)+ \left( \overline{y}_{U_{i_2:j_2}}-\tilde{y}_{s_{i_2:j_2}}\right)$.
		
		To conclude the proof, we just need to show that $a_{1N},a_{2N},a_{3N},a_{4N}$ converge in probability to zero as $N\rightarrow \infty$. The Markov inequality guarantees that $a_{1N}$ converges in probability to zero if its second moment does. Such moment can be written as
		\begin{align*}
		& \mathbb{E}(a_{1N}^2)\\
		&= \frac{n_N^2}{N_{i_1:j_1}^2 N_{i_2:j_2}^2} \underset{p,k \in U_{i_1:j_1} \cap U_{i_2:j_2} }{\sum} \frac{1-\pi_p}{\pi_p} \frac{1-\pi_k}{\pi_k} \left(y_p-\overline{y}_{U_{i_1:j_1}}\right)^2 \left(y_k-\overline{y}_{U_{i_2:j_2}}\right)^2 \frac{\Delta_{pk}}{\pi_p \pi_k}\\
		&+ \frac{2 n_N^2}{N_{i_1:j_1}^2 N_{i_2:j_2}^2} \underset{p \in U_{i_1:j_1} \cap U_{i_2:j_2} }{\sum} \underset{\underset{k\neq l}{k \in U_{i_1:j_1}, \;  l \in U_{i_2:j_2}} }{\sum} (y_p-\overline{y}_{U_{i_1:j_1}})(y_p-\overline{y}_{U_{i_2:j_2}})(y_k-\overline{y}_{U_{i_1:j_1}})(y_l-\overline{y}_{U_{i_2:j_2}}) \\
		&\times \frac{1-\pi_p}{\pi_p}\frac{\Delta_{kl}}{\pi_k \pi_l}\mathbb{E}\left(\frac{I_p-\pi_p}{\pi_p}\frac{I_kI_l-\pi_{kl}}{\pi_{kl}}\right)+ \frac{n_N^2}{N_{i_1:j_1}^2 N_{i_2:j_2}^2} \underset{\underset{p\neq q}{p \in U_{i_1:j_1}, \;  q \in U_{i_2:j_2}} }{\sum} \underset{\underset{k\neq l}{k \in U_{i_1:j_1}, \;  l \in U_{i_2:j_2}} }{\sum} \frac{\Delta_{pq}}{\pi_p \pi_q} \frac{\Delta_{kl}}{\pi_k \pi_l} \\
		& \times (y_p-\overline{y}_{U_{i_1:j_1}})(y_q-\overline{y}_{U_{i_2:j_2}})(y_k-\overline{y}_{U_{i_1:j_1}})(y_l-\overline{y}_{U_{i_2:j_2}}) \mathbb{E}\left( \frac{I_pI_q-\pi_{pq}}{\pi_{pq}} \frac{I_kI_l-\pi_{kl}}{\pi_{kl}}\right)\\
		&=b_{1N}+b_{2N}+b_{3N}.
		\end{align*} 
		Furthermore, 
		
		\begin{align*}
		|b_{1N}| &\leq \frac{n_N^2}{N^3 \lambda^3} \frac{N^4}{N_{i_1:j_1}^2N_{i_2:j_2}^2}  \left( \frac{\underset{k \in U_{i_1:j_1} \cap U_{i_2:j_2} }{\sum} \left(y_k-\overline{y}_{U_{i_1:j_1}}\right)^4}{N} + \frac{\underset{k \in U_{i_1:j_1} \cap U_{i_2:j_2} }{\sum} \left(y_k-\overline{y}_{U_{i_2:j_2}}\right)^4}{N} \right)\\
		&+ \frac{n_N^2 \underset{p,k \in U_N: p \neq k}{\max}|\Delta_{pk}|}{N^2 \lambda^4}  \frac{N^4}{N_{i_1:j_1}^2N_{i_2:j_2}^2} \\
		&\times \left( \frac{\underset{p \in U_{i_1:j_1} \cap U_{i_2:j_2} }{\sum} \left(y_p-\overline{y}_{U_{i_1:j_1}}\right)^4}{N} + \frac{\underset{k \in U_{i_1:j_1} \cap U_{i_2:j_2} }{\sum} \left(y_k-\overline{y}_{U_{i_2:j_2}}\right)^4}{N} \right)\\
		&\leq \frac{N^4}{N_{i_1:j_1}^2 N_{i_2:j_2}^2} \frac{n_N}{N \lambda^3} \left(\frac{n_N}{N^2}+ \frac{n_N \underset{p,k \in U_N: p \neq k}{\max}|\Delta_{pk}|}{N \lambda} \right) \\
		&\times \left( \frac{\underset{p \in U_{i_1:j_1} \cap U_{i_2:j_2} }{\sum} \left(y_p-\overline{y}_{U_{i_1:j_1}}\right)^4}{N} + \frac{\underset{k \in U_{i_1:j_1} \cap U_{i_2:j_2} }{\sum} \left(y_k-\overline{y}_{U_{i_2:j_2}}\right)^4}{N} \right)
		\end{align*}
		which converges to zero as $N \rightarrow \infty$ by Assumptions (A2)-(A5). Also, after separating the double sum in $b_{3N}$ into two sums where $(p,q)=(k,l)$ and $(p,q) \neq (k,l)$, we get that
		
		\begin{align*}
		|b_{3N}| &\leq O \left( \frac{1}{N} \right) + \frac{(n_N \underset{p,q \in U_N: p \neq q}{\max}|\Delta_{pq}|)^2}{\lambda^4 \lambda^{*2}} \frac{N^4}{N_{i_1:j_1}^2N_{i_2:j_2}^2}\underset{(p,q,k,l) \in D_{4,N}}{\max}|E[(I_p I_q-\pi_{pq})(I_k I_l-\pi_{kl})]|\\
		&\times \left( \frac{\underset{p \in U_{i_1:j_1} }{\sum} \left(y_p-\overline{y}_{U_{i_1:j_1}}\right)^4}{N} + \frac{\underset{q \in U_{i_2:j_2} }{\sum} \left(y_q-\overline{y}_{U_{i_2:j_2}}\right)^4}{N} \right)
		\end{align*}
		where the last term goes to zero by Assumptions (A2)-(A6). In addition, an application of the Cauchy-Schwarz inequality along with the fact that both $b_{1N}, b_{3N}$ tend to zero, shows that $b_{2N}$ converges to zero. Therefore, the Markov-inequality let us conclude that $a_{1N}=o_p(1)$. 
		
		Now, note that
		\begin{align*}
		a_{4N} &\leq \frac{N^2}{N_{i_1:j_1}N_{i_2:j_2}}|\tilde{y}_{s_{i_1:j_1}}-\overline{y}_{U_{i_1:j_1}}| |\tilde{y}_{s_{i_2:j_2}}-\overline{y}_{U_{i_2:j_2}}| \left( \frac{n_N}{N \lambda} + \frac{n_N \underset{k,l \in U_N: k \neq l}{\max}|\Delta_{kl}|)}{\lambda^2 \lambda^*} \right). 
		\end{align*}
		Then, $a_{4n}=o_p(1)$ since $\tilde{y}_{s_{i_1:j_1}}-\overline{y}_{U_{i_1:j_1}}=O_p(n^{-1/2})$ and $\tilde{y}_{s_{i_2:j_2}}-\overline{y}_{U_{i_2:j_2}}=O_p(n^{-1/2})$. Analogously, $a_{2N}=o_p(1)$ and $a_{3N}=o_p(1)$. Thus, 
		
		\begin{align*}
		n_N\left(\frac{\widehat{N}_{i_1:j_1}\widehat{N}_{i_2:j_2}}{N_{i_1:j_1}N_{i_2:j_2}}\widehat{AC}(\tilde{y}_{s_{i_1:j_1}},\tilde{y}_{s_{i_2:j_2}}) - AC(\tilde{y}_{s_{i_1:j_1}},\tilde{y}_{s_{i_2:j_2}})\right)=o_p(1).
		\end{align*}
		
		Finally, we have that $\frac{\widehat{N}_{i_1:j_1}\widehat{N}_{i_2:j_2}}{N_{i_1:j_1}N_{i_2:j_2}}-1=O_p(n^{-1/2})$ since $\frac{\widehat{N}_{i_1:j_1}}{N_{i_1:j_1}}-1=O_p(n^{-1/2})$ and $\frac{\widehat{N}_{i_2:j_2}}{N_{i_2:j_2}}-1=O_p(n^{-1/2})$. Therefore,
		
		\begin{align*}
		n_N\left(\widehat{AC}(\tilde{y}_{s_{i_1:j_1}},\tilde{y}_{s_{i_2:j_2}}) - \frac{\widehat{N}_{i_1:j_1}\widehat{N}_{i_2:j_2}}{N_{i_1:j_1}N_{i_2:j_2}}\widehat{AC}(\tilde{y}_{s_{i_1:j_1}},\tilde{y}_{s_{i_2:j_2}})\right)=o_p(1),
		\end{align*}
		which concludes the proof.
	\end{proof}

	\begin{proof}[Proof of Theorem 4]
		Fix $d$. First, recall that
		$$\tilde{\theta}_{s_d}=\underset{i\leq d}{\max} \; \underset{d\leq j}{\min} \; \tilde{y}_{s_{i:j}} \; \; \text{ and } \; \;  \theta_{U_d}=\underset{i\leq d}{\max} \; \underset{d\leq j}{\min} \; \overline{y}_{U_{i:j}}.$$
		By linearization arguments, it is true that $\tilde{y}_{s_{i:j}}-\overline{y}_{U_{i:j}}=O_p( n_N^{-1/2})$.
		
		Define $v_{s_i}=(\tilde{y}_{s_{i:d}},\tilde{y}_{s_{i:d+1}},\dots , \tilde{y}_{s_{i:D}} )^{\top}$ and $v_{U_i}=(\overline{y}_{U_{i:d}},\overline{y}_{U_{i:d+1}},\dots , \overline{y}_{U_{i:D}} )^{\top}$ for $i=1,2,\dots, d$. Hence, we have that
		$$v_{s_i}-v_{U_i}=\boldsymbol{1} O_p( n_N^{-1/2}). $$
		By Lemma 2, it is true that
		$$\min(v_{s_i}) -\min(v_{U_i})=O_p( n_N^{-1/2})$$
		Now, define $L_{s}=(\min(v_{s_1}),\min(v_{s_2}),\dots, \min(v_{s_d}))^{\top}$ and $L_{U}=(\min(v_{U_1}),\min(v_{U_2}),\dots, \min(v_{U_d}))^{\top}$. Therefore,
		$$L_{s}-L_{U}=\boldsymbol{1} O_p( n_N^{-1/2}).$$ 
		Finally, applying again Lemma 2 let us conclude that
		$$\max{L_s}-\max{L_U}=O_p( n_N^{-1/2}), $$
		which concludes the proof.
	\end{proof}

	
	\begin{proof}[Proof of Theorem 5]
		The CIC$_s$ difference between the constrained and the unconstrained estimator can be expressed as
		\begin{align*}
		\text{CIC}_s(\boldsymbol{\tilde{\theta}}_s) - \text{CIC}_s(\boldsymbol{\tilde{y}}_s) 
		&= (\boldsymbol{\tilde{y}}_{s}-\boldsymbol{\tilde{\theta}}_s)^{\top} \boldsymbol{W}_s (\boldsymbol{\tilde{y}}_{s}-\boldsymbol{\tilde{\theta}}_s) \\
		& -2 \Tr \left[\boldsymbol{W}_s \left( \widehat{\text{cov}}(\boldsymbol{\tilde{y}}_s, \boldsymbol{\tilde{y}}_s)- \widehat{\text{cov}}(\boldsymbol{\tilde{\theta}}_s, \boldsymbol{\tilde{y}}_s) \right)  \right] \\
		&= \delta_{1N} - 2\delta_{2N}. 
		\end{align*}
		
		First, assume that $\mu_1<\mu_2<\dots<\mu_D$. Define $A$ to the event where $\tilde{y}_{s_1}<\tilde{y}_{s_2}<\dots<\tilde{y}_{s_D}$, that is, $J_s^{0}=\emptyset$ and $J_{s}^{1}=\{1,2,\dots, D-1\}$. Then, from Theorem 1, we can conclude that $P(A^c)=o\left(1\right)$. Moreover, note that the CIC$_{s}$ difference is zero when $A$ holds. Hence,
		\begin{equation*}
		P\left( \text{CIC}_s(\boldsymbol{\tilde{y}}_s)<\text{CIC}_s(\boldsymbol{\tilde{\theta}}_s) \right) \leq P(A^c)=o(1).
		\end{equation*}
		Now, suppose that $\mu_1, \mu_2, \dots, \mu_D$ are not monotone. From Theorem 1 and Lemma 3, $\delta_{1N}-2\delta_{2N}=\left(\boldsymbol{\mu}-\boldsymbol{\theta_{\mu}}\right)^{\top}\boldsymbol{\Gamma}\left(\boldsymbol{\mu}-\boldsymbol{\theta_{\mu}}\right)+O_p(n_N^{-1/2})$. Further, $\left(\boldsymbol{\mu}-\boldsymbol{\theta_{\mu}}\right)^{\top}\boldsymbol{\Gamma}\left(\boldsymbol{\mu}-\boldsymbol{\theta_{\mu}}\right)>0$, since $\mu_1, \mu_2, \dots, \mu_D$ are not monotone. Thus,
		\begin{align*}
		P\left( \text{CIC}_s(\boldsymbol{\tilde{y}}_s) \geq \text{CIC}_s(\boldsymbol{\tilde{\theta}}_s)\right)=P(2\delta_{2N} \geq \delta_{1N})=o(1)
		\end{align*} 
		which concludes the proof.
	\end{proof}

	\begin{proof}[Proof of Theorem 6]
		We can write the PSE difference as
		\begin{align*}
		\text{PSE}(\boldsymbol{\widehat{\theta}}_s)- \text{PSE}(\boldsymbol{\widehat{y}}_s) &= [\mathbb{E}(\boldsymbol{\widehat{y}}_s)-\mathbb{E}(\boldsymbol{\widehat{\theta}}_s)]^{\top} \boldsymbol{W}_U [\mathbb{E}(\boldsymbol{\widehat{y}}_s)-\mathbb{E}(\boldsymbol{\widehat{\theta}}_s)] + 2\Tr\{\boldsymbol{W}_U[\text{var}(\boldsymbol{\widehat{y}}_s)-\text{var}(\boldsymbol{\widehat{\theta}}_s)]\}\\
		&= A_N + B_N .
		\end{align*}
		Assume first that $\mu_1 < \mu_2 < \dots < \mu_D$. This implies that $J_{\mu}^0=\emptyset$ and $J_{\mu}^1=\{1,2,\dots, D-1\}$ i.e. all points of the GCM are corner points. Based on the proof of Theorem 2 (with $\boldsymbol{P}_{F_1}=\boldsymbol{I}_D$), we have that $\mathbb{E}(\boldsymbol{\widehat{\theta}}_s) = \mathbb{E}(\boldsymbol{\widehat{y}}_s)+\boldsymbol{1}o(n_N^{-1})$ and $\text{var}(\boldsymbol{\widehat{\theta}}_s) = \text{var}(\boldsymbol{\widehat{y}}_s)+\boldsymbol{J}o(n_N^{-1})$. Therefore, $A_N=o(n_N^{-1})$ and $B_N=o({n_N^{-1}})$, which concludes the first part of the proof.
		
		Assume now that $\mu_1, \mu_2, \dots, \mu_D$ are not monotone. Lemma 5 and a direct application of Chebyshev's inequality imply that $\boldsymbol{\widehat{\theta}}_s-\mathbb{E}(\boldsymbol{\widehat{\theta}}_s)=\boldsymbol{1}O_p(n_N^{-1/2})$. Moreover, since $\boldsymbol{\widehat{\theta}}_s-\boldsymbol{\theta}_{\mu}=O_p(n_N^{-1/2})$, then $\mathbb{E}(\boldsymbol{\widehat{\theta}}_s)-\boldsymbol{\theta}_{\mu}=\boldsymbol{1}O(n_N^{-1/2})$. Hence, $A_N=(\boldsymbol{\mu}-\boldsymbol{\theta}_{\mu})^{\top}\boldsymbol{\Gamma}(\boldsymbol{\mu}-\boldsymbol{\theta}_{\mu}) + o(1)$, where the quadratic form is strictly greater than zero by the non-monotone assumption on the $\mu$'s. On the other hand, since both $\text{var}(\boldsymbol{\widehat{y}}_s)$ and $\text{var}(\boldsymbol{\widehat{y}}_s)$ are of the order $O(n_N^{-1/2})$, then $B_N=O(n_N^{-1})$. This concludes the proof.
	\end{proof}

\end{document}